\documentclass{aa} 
\usepackage[utf8]{inputenc}
\usepackage{graphicx}
\usepackage{txfonts}
\usepackage{lineno}
\linenumbers

\usepackage[draft=false]{hyperref}
\usepackage{natbib}
\usepackage{mathrsfs}
\usepackage[T1]{fontenc}
\usepackage{url}
\usepackage{gensymb}
\usepackage{verbatim}
\usepackage{multirow}
\usepackage{booktabs} 
\usepackage{orcidlink}
\usepackage{upgreek}
\usepackage{float}
\hypersetup{colorlinks,linkcolor={blue},citecolor={blue},urlcolor={blue}}

\newcommand{\dmunit}{pc\,cm$^{-3}$}

\begin{document}

\title{PSR J0024$-$7204ai: a massive, eccentric binary system in the globular cluster 47 Tucanae}
\titlerunning{a massive, eccentric binary system in the globular cluster 47 Tucanae}
\authorrunning{Risbud et al.}
\author{D. Risbud\orcidlink{0000-0001-9174-2883}\inst{1,2, 3}\thanks{\email{s14drisb@uni-bonn.de}},
A. Ridolfi\orcidlink{0000-0001-6762-2638}\inst{3}\thanks{\email{alessandro.ridolfi@uni-bielefeld.de}}, 
P. C. C. Freire\orcidlink{0000-0003-1307-9435}\inst{1}, 
M. Cadelano\orcidlink{0000-0002-5038-3914} \inst{4,5},
W. Chen\orcidlink{0000-0002-6089-7943}\inst{1},
L. Zhang\orcidlink{0000-0001-8539-4237}\inst{10,11},
R. Nag\orcidlink{0009-0005-6754-0655}\inst{6},
F. Camilo\orcidlink{0000-0002-1873-3718}\inst{7},
P.V. Padmanabh\orcidlink{0000-0001-5624-4635}\inst{8,9},
A. Corongiu\orcidlink{0000-0002-5924-3141}\inst{6},
F. Abbate\orcidlink{0000-0002-9791-7661}\inst{6},
A. Possenti\orcidlink{0000-0001-5902-3731}\inst{6}
}
\institute{
Max-Planck-Institut für Radioastronomie, Auf dem H\"{u}gel 69, D-53121 Bonn, Germany
\and
Rheinische Friedrich-Wilhelms-Universität Bonn, Regina-Pacis-Weg 3, D-53113 Bonn, Germany
\and
Fakultät für Physik, Universität Bielefeld, D-33615 Bielefeld, Germany
\and
Dipartimento di Fisica e Astronomia, Università degli Studi di Bologna, Via Gobetti 93/2, I-40129 Bologna, Italy
4
\and
INAF, Osservatorio di Astrofisica e Scienza dello Spazio di Bologna, Via Gobetti 93/3, I-40129 Bologna, Italy
\and
INAF – Osservatorio Astronomico di Cagliari, Via della Scienza 5, I-09047 Selargius (CA), Italy
\and
South African Radio Astronomy Observatory, 2 Fir Street, Black River Park, Observatory 7925, South Africa
\and
Max Planck Institute for Gravitational Physics (Albert Einstein Institute), D-30167 Hannover, Germany
\and
Leibniz Universit\"{a}t Hannover, D-30167 Hannover, Germany
\and
National Astronomical Observatories, Chinese Academy of Sciences, A20 Datun Road, Chaoyang District, Beijing 100101, People's Republic of China
\and
Centre for Astrophysics and Supercomputing, Swinburne University of Technology, P.O. Box 218, Hawthorn, VIC 3122, Australia
}

\abstract{In this paper we present PSR J0024$-$7204ai, a 13.026-ms binary pulsar recently discovered in the globular cluster 47 Tucanae by the MeerKAT radio telescope.  This is the slowest spinning pulsar known in this globular cluster, and has a $\sim1.67$-day orbit with an eccentricity of $e\approx0.18$. Although it was not yet possible to derive an unambiguous phase-connected timing solution, by combining detections obtained from MeerKAT and archival Parkes data we were able to measure the rate of advance of periastron to high significance, $\dot{\omega}$ =  0.1601 $\pm 0.0046$ deg yr$^{-1}$.
This value implies a total system mass of $2.41 \pm 0.11\, \mathrm{M}_\odot$ (68.3\% C. L.), which, when combined with the binary mass function, gives a maximum pulsar mass of $\sim 1.7 \, \mathrm{M}_\odot$ and a minimum companion mass of $\sim 0.7\, \mathrm{M}_\odot$. Apart from being the slowest pulsar in 47~Tucanae, its orbit is by far the most eccentric and its companion is the most massive among all known binary pulsars in this globular cluster. 
One possibility is that system is an old MSP - Carbon-Oxygen White Dwarf binary, whose orbit was perturbed by stellar dynamical interactions in the cluster core. Further follow-up observations of this system will be essential for a more detailed characterisation of this system and its evolution.}

\maketitle

\section{Introduction}

Globular clusters (GCs) are gravitationally self-bound Galactic subsystems consisting of 10$^4$-10$^6$ stars.
They are also among the oldest stellar systems in the Galaxy, with their ages
ranging from 10 to 13 billion years—comparable to the age of the Universe.
In their cores, the stellar density can be as high as 10$^5$-10$^6$ M$_\odot$ pc$^{-3}$ \citep{2018MNRAS.478.1520Baumgardt}, which are a few orders of magnitude higher than what is typically seen in the Galactic field.
These extremely dense environments enhance two or three-body stellar dynamical
interactions and formation of the binary stellar systems \citep{1975AJ.....80..809Hills,1987IAUS..125..187Verbunt, 1995ApJS...99..609Sigurdsson}.

In such exchange encounters, old and ``dead'' neutron stars (NSs) (the ones that have crossed the death line, and their emission is no more detectable) can become members of binary systems with low-mass main sequence stellar companions. As the companion star evolves according to the standard theory of stellar evolution, and fills its Roche lobe, the mass and angular momentum is transferred to the NS. This accretion of matter onto the NS surface makes the pair visible in X-ray wavelengths as a low-mass X-ray binary (LMXB). The LMXBs per unit of stellar mass are orders of magnitude more numerous in GCs than in the Galactic disk \citep{1975ApJ...199L.143Clark}.

Due to the relatively slower evolution of the low-mass companion, the LMXB phase is long-lived and can spin up the NS to rotation periods of just  a few milliseconds, giving birth to a radio millisecond pulsar (MSP). This process is known as \emph{pulsar recycling} (\citealt{1982CSci...51.1096Radhakrishnan, 1982Natur.300..728Alpar, 1991PhR...203....1Bhattacharya,  2013Natur.501..517Papitto}, for a review see \citealt{2023pbse.book.....Tauris}).  The large rate of formation of LMXBs in GCs is the reason why a wealth of MSPs ($P < 10$~ms) are seen in these environments\footnote{For an updated list, see \url{https://www3.mpifr-bonn.mpg.de/staff/pfreire/GCpsr.html}.}. Such MSPs are formed with very low, near-zero eccentric orbits due to the tidal circularization.

 If the companion is a MS star significantly more massive than $1\, \rm M_\odot$, the systems become intermediate-mass and high-mass X-ray binaries, respectively. These have
relatively short-lived mass transfer phases, since the companion evolves faster. The pulsar is spun-up to spin periods between 10 to 200 ms, resulting in a mild recycling \citep{2006csxs.book..623Tauris}. The companion stars in these systems are either relatively massive CO WDs or ONeMg WDs, in which case the orbits have very low eccentricities, $e \lesssim 10^{-3}$ \citep{2000ApJ...530L..93Tauris,2011MNRAS.416.2130Tauris,2012MNRAS.425.1601Tauris} or in some cases NS \citep{2017ApJ...846..170Tauris},  in which case $e \sim 0.1$ and larger. However, it is important to point out that such massive MS companions were only available in the earlier stages of the evolution of this globular cluster.
 
The binary MSPs in globular clusters are particularly interesting because of stellar perturbations, which can make them acquire significant eccentricities  even if they're born circular. In some cases these stellar encounters lead to exchange interactions, where a low-mass companion to a an already recycled pulsar is exchanged by a more massive star, like M15C \citep{1991ApJ...374L..41Prince}, NGC 1851A, D and E \citep{2004ApJ...606L..53Freire,2022A&A...664A..27Ridolfi,2024Sci...383..275Barr}, NGC 6624G \citep{2021MNRAS.504.1407Ridolfi}, M30B \citep{2023ApJ...942L..35B} and NGC 6544B \citep{2012ApJ...745..109Lynch}.
These systems are therefore {\em secondary exchange encounters}.
Their large eccentricities and companion masses imply that with several years of timing data, relativistic Post-Keplerian (PK) parameters like the rate of advance of periastron ($\dot\omega$), the Einstein delay ($\gamma_{\rm E}$) or the two parameters characterising the Shapiro delay \citep{1964PhRvL..13..789Shapiro} become measurable, allowing a determination of the masses \citep{2012ApJ...745..109Lynch,2025A&A...697A.166Dutta} and, in some cases, like that of PSR B2127+11C, tests of general relativity \citep{2006ApJ...644L.113Jacoby}. However, this type of binaries is only detected in GCs where the rate of stellar interactions per binary ($\gamma$) is large \citep{2014A&A...561A..11Verbunt}.

NGC 104, or alternatively 47 Tucanae (47 Tuc here on), is located in the southern hemisphere at equatorial sky coordinates $\alpha = 00^\mathrm{h}\,24^\mathrm{m}\,05.67^\mathrm{s}$,  $\delta = -72^\circ\,04'\,52.6''$, or Galactic coordinates $l = 305.89^\circ$, $b = -44.89^\circ$ at a distance of 4.69 kpc from the Sun \citep{2012AJ....143...50Woodley}. Its age estimates vary from 10.4 to 13.4 Gyr (\citealt{2003A&A...408..Gratton,2017MNRAS.468..645Brogaard,2020MNRAS.492.Thompson}). Several past discoveries and timing studies of the pulsars in 47 Tuc \citep{1990Natur.345..598Manchester,1991Natur.352..219_Manchester,1995MNRAS.274..547Robinson,2000ApJ...535..975Camilo,2016MNRAS.462.2918Ridolfi,2017MNRAS.471..857Freire} enabled studies of cluster dynamics, mass segregation, gravitational potential and the detection of ionized gas in the cluster \citep{2001ApJ...557L.105Freire_2001c, 2018MNRAS.481..627Abbate,2023MNRAS.518.1642Abbate}.
The precise positions allowed X-ray detections of all pulsars with known timing solutions \citep{2002ApJ...581..470Grindlay,2005ApJ...630.1029Bogdanov,2006ApJ...646.1104Bogdanov,2017MNRAS.472.3706Bhattacharya,2021MNRAS.500.1139Hebbar}.
Of the 13 binaries with precise positions, five white dwarf \citep{2001ApJ...557L..57Edmonds, 2015MNRAS.453.Rivera,2015ApJ...812...63Cadelano} and one redback \citep{2002ApJ...579..741Edmonds} companions were detected and characterised in archival {\em Hubble Space Telescope} (HST) images.

This scientific payoff motivated MeerKAT observations of 47 Tuc, which began under the TRAPUM\footnote{\url{http://www.trapum.org}} project \citep{Stappers_Kramer2016} in May 2020. The main aim has been discovering new pulsars, but also measuring the masses of some of the previously known pulsars. At the time of writing, the project has discovered 17 pulsars in the cluster (\citealt{2021MNRAS.504.1407Ridolfi}; Chen et al., in prep. ). Of these new discoveries, at least 14 are  binary pulsars; the total percentage of binaries is now 29/42 = 69 per cent.

Unlike the binary systems in some other globular clusters (like Terzan 5; \citealt{2024A&A...686A.166_Padmanabh}), all binaries in 47~Tuc have low ($< 0.1$) orbital eccentricities and companion masses. This this expected for GCs like 47 Tuc that has a relatively low value of $\gamma$ ; i.e., once a low-mass X-ray binary forms, it evolves into a circular MSP - low mass companion system without much risk of being perturbed, at least for the systems with smaller orbital periods, where the cross section for interactions is smaller.

One of the newly discovered pulsars, PSR~J0024$-$7204ai (henceforth to be known as 47~Tuc~ai) is the exception among the binaries in 47 Tuc, and is the topic of this work. As described below, its spin period of 13.026 ms   
makes it the first mildly recycled pulsar in 47 Tuc. Its eccentric binary orbit ($e\, \simeq 0.18$) and large companion mass also make it stand out from the remaining pulsar population.
The structure of the paper is as follows: In Sec. \ref{sec:data}, we give an overview of the dataset used for this work. Sec. \ref{sec:Results} focuses on the determination of the orbit, localization and basic emission characteristics, and an early timing analysis. In Sec. \ref{sec:nature_and_origin} we discuss the nature and origin of this binary system. In Sec. \ref{sec:conclusion}, we summarise our results and discuss some near-future prospects.  

\section{Observations}\label{sec:data}

\subsection{MeerKAT observations}

47 Tuc was observed with the South African MeerKAT radio telescope array on 41 occasions between 2019 March and 2024 January. The observations were made as part of the TRAPUM  and MeerTime\footnote{\url{http://www.meertime.org}} \citep{Bailes+2020} Large Survey Projects and had a number of different scientific aims. Therefore the set-ups and observing parameters of the whole dataset are highly heterogeneous.
Observations carried out under MeerTime, which had pulsar timing as the main scientific driver, made use of the Pulsar Timing User Supplied Equipment (PTUSE) as the primary backend. The latter is capable of recording up to four tied-array beams on the sky and record the data as coherently de-dispersed search-mode PSRFITS files with full-Stokes information. Given the small number of beams that could be synthesized, the PTUSE beams were produced correlating only the central 44, 1-km-core antennas of the MeerKAT array, so as to have a larger field of view at the cost of $\sim30$ per cent reduction in raw sensitivity.
Observations carried out under TRAPUM, whose main goal was the discovery of several new pulsars, made use of the Filterbanking Beamformer User Supplied Equipment (FBFUSE) and the Accelerated Pulsar Search User Supplied Equipment (APSUSE) as the primary backend. These enabled the synthesis of up to 288 tied-array beams on the sky, which in turned allowed the coverage of a much larger sky area around the cluster center compared to PTUSE, even correlating the raw signals coming from all the available (up to 64) MeerKAT antennas, hence retaining maximum sensitivity. In this case, each beam could be recorded as a search-mode ``filterbank'' file, in total-intensity only and no coherent de-dispersion. However, since the DM of 47 Tuc is well known, the TRAPUM beams were typically recorded with 4096 frequency channels, and later incoherently de-dispersed at a DM of $24.4$ \dmunit\ and summed in groups of 16, so as to reduce the number of channels to 256.
In the vast majority of the observations, both PTUSE and FBFUSE+APSUSE were used in parallel.
 All the observations were made with either the UHF-band (544-1088 MHz) or the L-band (856-1712 MHz) receivers.
The details of all the observations used in this work are reported in Table \ref{table:observations}.

\begin{table*}%[!htbp]
\caption{Observations of 47 Tuc made by MeerKAT under MeerTime and TRAPUM projects.}
\centering      
\begin{footnotesize}
\begin{tabular}{lcccrccllcc}
\hline
\hline
Obs   & Start Date and Time   & Length    & Backend &  $t_{\rm samp}$  & $f_{\rm c}$   & $\Delta f$        &  $N_{\rm pol}$        & $N_{\rm b}$ & $N_{\rm ant}$ & (S/N) / $N_{\rm ToAs}$ \\
ID    &  (UTC)                & (min)   &  &  ($\upmu$s)        & (MHz)         & (MHz)             &         &      &            &  \\
\hline
01L  & 2019-03-15 09:16 & 16  & PTUSE & 76.56  & 1176 & 214            & 1 &            1         & 59           &       -     \\
01L  & 2019-03-15 09:16 & 16  & PTUSE & 76.56  & 1390 & 214            & 1 &            1         & 59           &       -     \\
02L  & 2019-10-09 22:02 & 60  & PTUSE & 76.56  & 1284 & 642            & 1 &            1         & 40           &       -     \\
03L  & 2019-10-19 01:49 & 150 & PTUSE & 76.56  & 1284 & 642            & 1 &            1         & 39           &       -     \\
04L  & 2019-11-05 21:15 & 90  & PTUSE & 76.56  & 1284 & 642            & 1 &            1         & 59           &       -     \\
05L  & 2019-12-14 17:04 & 60  & PTUSE & 76.56  & 1284 & 642            & 1 &            1         & 40           &       -     \\
06L  & 2019-12-19 18:47 & 60  & PTUSE & 76.56  & 1284 & 642            & 1 &            1         & 42           &       -     \\
07L  & 2020-01-01 08:23 & 60  & PTUSE & 76.56  & 1284 & 642            & 1 &            1         & 41           &       -     \\
08L  & 2020-01-27 21:47 & 60  & PTUSE & 76.56  & 1284 & 642            & 1 &            1         & 38           &       -     \\
09U  & 2020-03-20 17:43 & 90  & PTUSE & 75.29  &  816 & 544            & 1 &            1         & 38           &       -    \\
10L  & 2020-05-02 06:05 & 239 & APSUSE  & 76.56  & 1284 & 856            & 1 &            279         &   56         &     8.7 / 1       \\
11L  & 2020-07-29 04:07 & 50  & APSUSE  & 76.56  & 1284 & 856            & 1 &            289         &   56       &   20.3 / 5  \\
12U  & 2020-11-10 11:42 & 52  & APSUSE  & 60.24  &  816 & 544            & 1 &            277         &  56          &   8.6 / 2  \\
14L  & 2020-11-21 18:38 & 59  & APSUSE  & 76.56  & 1284 & 856            & 1 &            289         &  56          &   -         \\
15L  & 2020-12-17 11:23 & 49  & APSUSE  & 76.56  & 1284 & 856            & 1 &            289         &   56       &   9.5 / 2    \\
17L  & 2021-01-14 12:50 & 48  & APSUSE  & 76.56  & 1284 & 856            & 1 &            289         &   56         &   -         \\
18U  & 2021-01-22 11:58 & 59  & APSUSE  & 60.24  &  816 & 544            & 1 &            289         &   56         &   18.3 / 3   \\
19U  & 2021-01-30 12:54 & 59  & APSUSE  & 60.24  &  816 & 544            & 1 &            278         &    56        &  -         \\
20U  & 2021-02-06 08:49 & 59  & APSUSE  & 60.24  &  816 & 544            & 1 &            278         &   56         &  8.4 / 1      \\
21U  & 2021-02-17 04:53 & 59  & APSUSE  & 60.24  &  816 & 544            & 1 &            278         &  56          & 4.4 / 1     \\
22U1 & 2021-06-26 03:03 & 239 & PTUSE & 37.65  &  816 & 544            & 4 &            2          &  31          &   8.2 / 1        \\
22L1 & 2021-06-26 03:03 & 239 & PTUSE & 37.65  &  1284 & 856            & 4 &            2          &  31          &    -        \\
22U2 & 2021-06-26 07:16 & 179 & PTUSE & 37.65  &  816 & 544            & 4 &            1          &  31          &   11.0 / 1        \\
22L2 & 2021-06-26 07:16 & 179 & PTUSE & 37.65  &  1284 & 856            & 4 &            1          &  31          &   11.3 / 1        \\
23U  & 2021-07-29 18:38 & 120 & PTUSE  & 75.29  &  816 & 544            & 1 &            1         &  41          &    20.3 / 4    \\
25U  & 2022-01-26 15:32 & 119 & APSUSE  & 60.25  &  816 & 544            & 1 &            94         &     60     &  17.2  / 3        \\
26U1 & 2022-01-27 05:10 & 116 & APSUSE  & 60.25  &  816 & 544            & 1 &            94         &     56     &  39.0  / 8     \\
26U2 & 2022-01-27 07:22 & 299 & APSUSE  & 60.25  &  816 & 544            & 1 &            94         &      56    &  20.7 / 5     \\
26U3 & 2022-01-27 14:09 & 55  & APSUSE  & 60.24  &  816 & 544            & 1 &            94         &      56    &  9.6 / 2     \\
26U4 & 2022-01-27 15:18 & 179 & APSUSE  & 60.24  &  816 & 544            & 1 &            94         &      56    & 9.1  / 1          \\
26U5 & 2022-01-27 18:32 & 209 & APSUSE  & 60.24  &  816 & 544            & 1 &            96         &      56  &    9.8  / 1    \\
26U6 & 2022-01-27 22:10 & 58  & APSUSE  & 60.24  &  816 & 544            & 1 &            94         &      56    &  -     \\
27U  & 2022-01-28 05:10 & 114 & APSUSE  & 60.24  &  816 & 544            & 1 &            94         &      60      & 10.2 / 1 \\
28U  & 2022-01-29 07:08 & 102 & APSUSE  & 60.24  &  816 & 544            & 1 &            278         &    60      & 11.6 / 1          \\
29U  & 2022-06-09 03:07 & 119 & APSUSE  & 60.24  &  816 & 544            & 1 &            2         &    56        &  -          \\
30U  & 2022-11-11 23:10 & 120 & PTUSE & 60.24  &  816 & 544            & 4 &            2         & 51           &       11.3 / 1 \\
31U  & 2023-05-26 22:54 & 119 & APSUSE  & 60.24  &  816 & 544            & 1 &           2          &    56        &  -          \\
32U  & 2023-08-16 17:41 & 121 & APSUSE  & 120.47 &  816 & 544            & 1 &           2          &    56        &  -          \\
33U  & 2023-10-25 15:22	& 119 & APSUSE & 120.47  & 816 & 544              & 1 &           2          & 52           &   13.3 / 2        \\
34U  & 2023-12-13 16:07 & 120 & PTUSE & 60.24  &  816 & 544            & 1 &            2        & 51           &       -     \\
35U  & 2024-01-25 12:52 & 120 & PTUSE & 60.24  &  816 & 544            & 4 &            2          & 47           &      10.2  / 2    \\
36U  & 2024-02-11 10:52 & 120 & PTUSE & 60.24  &  816 & 544            & 4 &            2          & 53           &        10.6 / 2   \\
37U  & 2024-05-16 03:24 & 120 & PTUSE & 60.24  &  816 & 544            & 4 &            1          & 53           &         12.0  / 2 \\
38U  & 2024-05-20 09:47 & 120 & PTUSE & 60.24  &  816 & 544            & 4 &            2          & 49           &      -      \\
39U  & 2024-10-15 16:19 & 40  & PTUSE & 60.24  &  816 & 544            & 4 &            3          & 50           &      -      \\
40U  & 2024-11-23 00:59 & 120 & PTUSE & 60.24  &  816 & 544            & 4 &            3         & 50           &       -     \\
41U  & 2024-12-03 20:06 & 119 & PTUSE & 60.24  &  816 & 544            & 4 &            3          & 50       &        - \\
\hline
\end{tabular}
\end{footnotesize}
\tablefoot{ $t_{\rm samp}$: sampling time;   $f_{\rm c}$: central frequency;   $\Delta f$: nominal observing bandwidth;          $N_{\rm pol}$: number of recorded Stokes parameters;        $N_{\rm b}$: number of recorded tied-array beams; $N_{\rm ant}$: number of MeerKAT antennas used; S/N: signal-to-noise ratios of the detections, calculated from  \texttt{pdmp} routine of \texttt{PSRCHIVE} from the cleaned folded archives keeping the sub-integrations and frequency channels with pulsar signal. The observations in which the pulsar was not detected are marked with "-". $N_{\rm ToAs}$: number of ToAs that were extracted from that detection.  Observations with IDs 13L and 16L failed and did not yield any useful data. Observation 24U (made on 20 Aug 2021) used a single beam pointed at 47 Tuc H that did not cover the position of pulsar 47 Tuc ai.}
\label{table:observations}
\end{table*}

\subsection{Parkes archival data}

Given the highly negative $- 72 \deg$ declination of the GC 47 Tuc, the latter can only be observed by the radio telescopes in the southern hemisphere. Before the MeerKAT radio telescope became operational, it was possible to observe this GC only with the Parkes radio telescope in Australia. The latter carried out observations of 47 Tuc regularly from 1997 to 2013  with the Multibeam receiver \citep{Staveley-Smith+1996}, for a total of approximately 1770 observing hours, split across 414 observation epochs (i.e. days) and 519 total pointings\footnote{The logs of all 519 pointings can be found at \url{https://www3.mpifr-bonn.mpg.de/staff/pfreire/47Tuc/Observations_Table.html}.  }. For the detailed description of the dataset and its previous uses, refer to \citet{2016MNRAS.462.2918Ridolfi}. 

From the year 2019 onwards, the Parkes radio telescope was equipped with the ``Ultra-Wide bandwidth Low receiver'' (UWL) \citep{2020PASA...37...12Hobbs}, capable of observing in the frequency range 704-4032 MHz. A few observations of 47 Tuc were made with this receiver in the context of different projects (see Table \ref{table:Parkes_UWL_obs}) and were available on the CSIRO Data Access Portal. These observations were also processed in the context of this work.

\begin{table}
    \centering
    \caption{Observations of 47 Tuc made by the Parkes radio telescope with UWL frontend.}
    \begin{footnotesize}
    \begin{tabular}{ccccc}
        \hline \hline
        Project /   &  Start date   & Length    &   $t_{\rm samp}$ & DM \\
       
        Proposal ID    & and time (UTC)   & (min)   &    ($\upmu$s) & (pc cm$^{-3})$  \\
        \hline
        P981  &   2018-11-16 12:39 & 60& 64  & 24.6 \\
        P982  &   2019-01-04 07:06 & 120& 64 & 24.4 \\
        P982  &   2019-01-04 09:08 & 120& 64 & -   \\
        P982  &   2019-01-04 11:16 & 120& 8 & 24.4 \\
        P1006 &  2019-08-19 15:32 & 440& 64 & 24.36 \\

        P1022 &  	2019-11-18 08:54 & 481 & 64 & 24.36 \\
        P1022 &  	2019-11-23 09:06 & 293 & 64 & 24.36 \\
        P1022 &  	2019-11-23 09:06 & 293 & 64 & 24.36 \\
        P1022 &    2019-11-24 14:14 & 24 & 64 & 24.36 \\
        P1022 &    2019-11-24 15:19 & 24 & 64 & 24.36 \\
        P1022 &    2019-11-25 10:27 & 63 & 64 & 24.36 \\
        P1022 &    2019-11-26 09:50 & 127 & 64 & 24.36 \\
        P1022 &    2019-12-01 11:16 & 105 & 64 & 24.36 \\
        P1022 &    2019-12-02 09:44 & 84 & 64 & 24.36 \\
        P1022 &    2019-12-08 08:44 & 385 & 64 & 24.36 \\
        P1022 &    2019-12-08 08:44 & 385 & 64 & 24.36 \\
        P1022 &    2019-12-11 12:24 & 97 & 64 & 24.36 \\
        P1022 &    	2019-12-14 13:21 & 69 & 64 & 24.36 \\
        P1022 &    	2019-12-16 13:20 & 99 & 64 & 24.36 \\
        P1022 &    	2019-12-17 13:17 & 104 & 64 & 24.36 \\
        P1022 &    	2019-12-18 13:20 & 98 & 64 & 24.36 \\
        P1022 &    	2019-12-19 13:19 & 95 & 64 & 24.36 \\
        P1022 &    	2019-12-20 12:49 & 122 & 64 & 24.36 \\	
        P1054 &   2020-04-14 01:20 & 100& 64 &  24.4\\
        PX061 &   	2020-05-20 20:52 & 37 & 64 &  -\\
        P1076 &    	2021-01-25 04:59 & 106 & 64 & 24.36\\
        P595 & 2022-09-07 10:01 & 90 & 64 & - \\ 
        PX145 & 2025-12-17 08:16 & 223 & 64 &24.4\\
        \hline
    \end{tabular}
    \end{footnotesize}
    \tablefoot{ $t_{\rm samp}$ is the sampling time and the DMs used for coherent dedispersion are listed in the last column. Pulsar 47 Tuc ai was not detected in any of these observations. The observation under P595 was done between 1344.5 to 1599.5 MHz frequency range and the rest have entire UWL bandwidth of 704 to 4032 MHz. }
    \label{table:Parkes_UWL_obs}
\end{table}

\section{Preliminary characterization}\label{sec:Results}
\subsection{Discovery and orbital parameters}\label{sec:disc_orb}

47 Tuc ai  was first detected in the MeerKAT observation ``11L'' as a 13.026-ms candidate with a barycentric line of sight acceleration of 4.02 m\,s$^{-2}$ and a DM = 24.47 \dmunit\ (we refer to Chen et al., in prep. for more details about the search).

\begin{figure}
    \includegraphics[width=1\linewidth, height =0.70\linewidth]{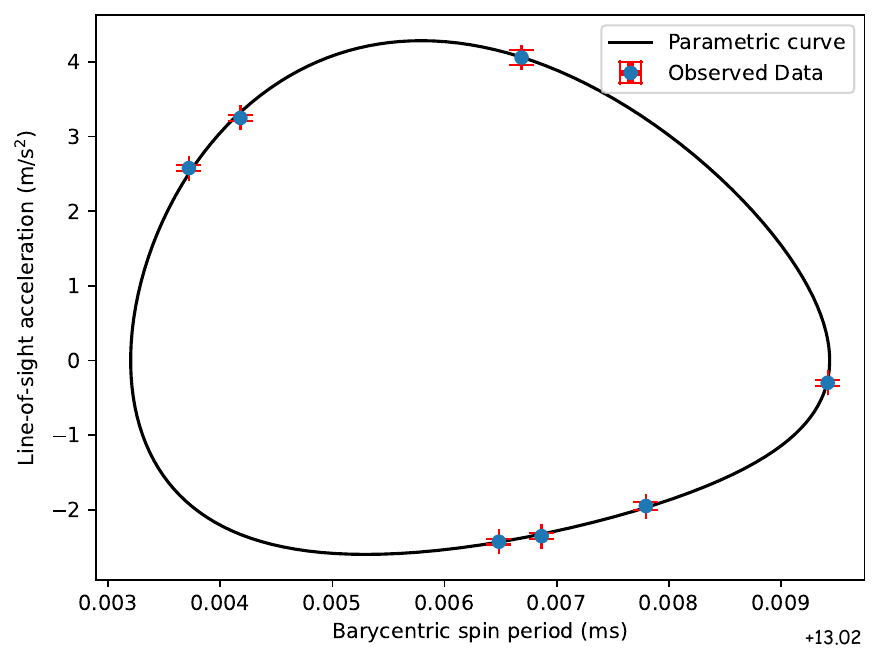}
    \caption{Period-Acceleration diagram for 47 Tuc ai. The  data points are the barycentric spin periods and line-of-sight accelerations as measured on seven different epochs where the pulsar was blindly detected. The best-fit parametric curve (solid black line), from equations 3 and 4 from \citet{2001MNRAS.326..901Freire_b}, largely deviates from a simple ellipse, implying a mild orbital eccentricity ($e \approx 0.18$). }
    \label{fig:P_A_curve}
\end{figure}

The large acceleration clearly pointed towards a binary nature. The candidate was soon confirmed to be a real pulsar based upon its detections in neighbouring beams of the 11L observation and by a few additional detections in other observations. In the first TRAPUM observing campaign on 47 Tuc, the pulsar was blindly detected in 7 observations, 2 of which were made at L-band and 5 others in the UHF-band. Plotting the measured spin periods vs the line-of-sight accelerations (Fig. \ref{fig:P_A_curve}) in a period-acceleration diagram (\citealt{2001MNRAS.322..885Freire2001a}, corrected as per the erratum \citealt{2009MNRAS.395.1775Freire_erratum}), revealed a mild orbital eccentricity and allowed a first estimate of the orbital parameters: the pulsar is in an orbit with an orbital period of $P_{\rm b} \simeq 1.65$~days, and a projected semi-major axis $x_{\rm p} \simeq 5.35$~s, an eccentricity of $e \simeq 0.18$, and a periastron longitude of $\omega \simeq 227$\textdegree.

After building a basic ephemeris with these parameters, we used the  \texttt{spider\_twister}\footnote{\url{https://github.com/alex88ridolfi/SPIDER_TWISTER}} software to find the precise times of passage at periastron ($T_0$) closest to each detection epoch.

These were used to apply the periodogram method (see e.g., \citealt{2016MNRAS.462.2918Ridolfi}), which allows a refinement of $P_{\rm b}$ by finding a value of the latter that fits an integer number of times between any two $T_0$ values. Finally, we used the \texttt{fitorb.py} code from the \texttt{PRESTO}\footnote{\url{https://github.com/scottransom/presto}} pulsar searching software \citep{Ransom2001} to fit directly the observed spin period of the pulsar as a function of time, which significantly improved the precision of the orbital parameters.
The refined parameters were used to update the pulsar ephemeris, which was in turn used to re-fold the whole Parkes and MeerKAT datasets. In doing so, we allowed for a full-orbit search in the value of $T_0$ with \texttt{spider\_twister}, to account for a possible wrong orbital phase prediction due to a possible still inaccurate $P_{\rm b}$. This led to the detection of 47 Tuc ai in a few additional MeerKAT observations, where the pulsar was not detected blindly, as well as in three different Parkes observations, dating back to the year 2000. These are listed in Table \ref{table:Parkes_multibeam_detections}. The three Parkes detections turned out to be very important for the precise measurement of the periastron advance (see Sec. \ref{sec:timing}). 

\begin{table}[h]
\centering
\caption{Observations of 47 Tuc made by the Parkes radio telescope with Multibeam receiver in which pulsar 47 Tuc ai was detected.}
\begin{footnotesize}
\begin{tabular}{cccc}
\hline \hline
Obs. ID   &  Start date   & Length & (S/N) / $N_{\rm ToAs}$ \\
       
          &                & (min) &     \\
\hline
        
47T040\_7  & 2000-02-17 03:24 & 220 & 2.9 / 2     \\
        
47T122\_1  & 2004-04-07 23:46	& 340 & 2.8 / 2    \\
        
47T196\_3  & 2013-08-04 17:41 & 223 & 4.8 / 2    \\      
\hline
\end{tabular}
\end{footnotesize}
\tablefoot{These observations were made at the central frequency of 1389.75 MHz, with 256 MHz bandwidth divided into 512 channels, at a sampling time of 80 $\upmu$s. All 519 observations can be found at \url{https://www3.mpifr-bonn.mpg.de/staff/pfreire/47Tuc/Observations_Table.html}. The $S/N$ was calculated in a same way as done for the cleaned folded archives from the MeerKAT observations.} 
\label{table:Parkes_multibeam_detections}
\end{table}

\subsection{\texttt{SEEKAT} Localization}\label{sec:localization}
\begin{figure}
    \includegraphics[width=1\linewidth, height =1\linewidth]{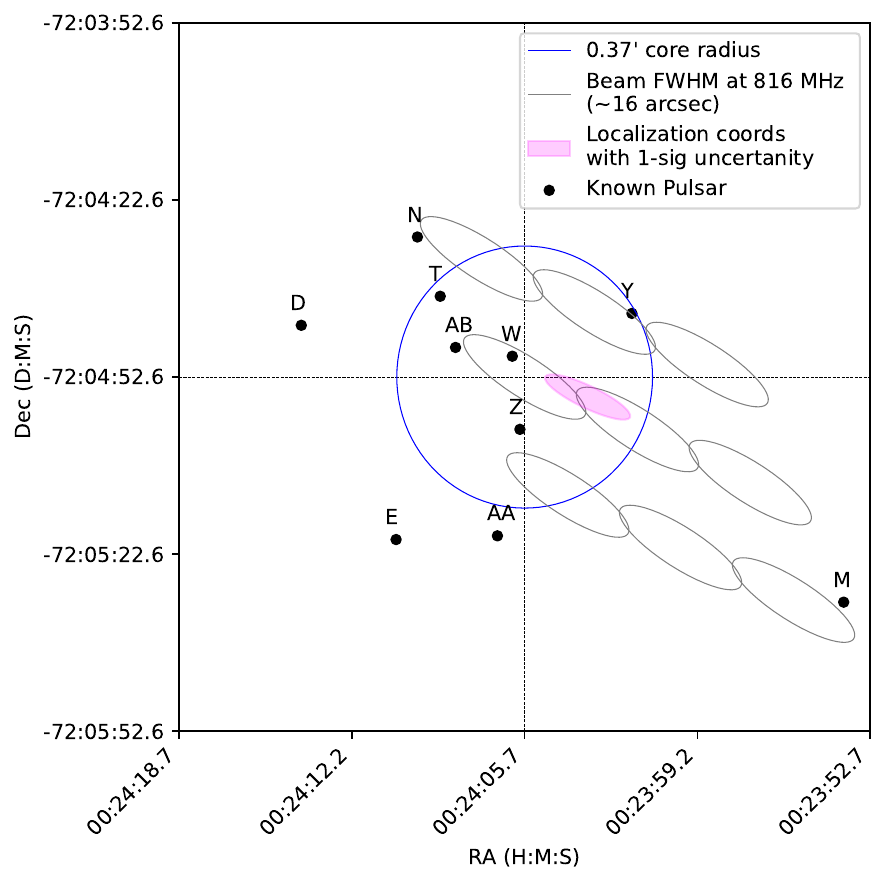}
    \caption{\textsc{SEEKAT} Localization of 47 Tuc ai, with the 1-$\sigma$ uncertainty on the localized position are shown as magenta coloured ellipse. The gray ellipses are beams from the observation 26U1 in which the pulsar was detected at the highest S/N, and were used for localizing the pulsar as discussed in sec. \ref{sec:localization}. The center of the GC 47 Tuc (central cross), the core radius (blue circle) and the positions of some of the known pulsars in 47 Tuc are shown for the reference. The image covers an area of 1 arcmin $\times$ 1 arcmin around the center of the GC $00^\mathrm{h}\,24^\mathrm{m}\,05.67^\mathrm{s}$, 
$ $-$72^\circ\,04'\,52.6''$ \citep{2012AJ....143...50Woodley}.  }
    \label{fig:Localization}
\end{figure}

MeerKAT observations under the TRAPUM project recorded up to 288 synthesized coherent beams in a dense tiling spanning a few arcminutes around the center of 47 Tuc. This unique capability of the TRAPUM recording system allows the determination of the sky position of newly discovered pulsars to arcsec precision, using just a single observation. The individual coherent beams have an average full-width at half maximum (FWHM) of $\approx$ 16 arcsec on the sky at UHF band (central frequency of 816 MHz). The detection of any pulsar in a given beam indicates that the true position of pulsar is within the beam FWHM or quite close to the beam bore-sight. Given that a pulsar has detections in multiple surrounding beams, this information can be exploited to find the true position with higher accuracy. We used the  \texttt{SEEKAT} software \citep{2023RASTI...2..114Bezuidenhout}, which takes the positions and point spread function of multiple coherent beams, and the detection S/N values in those beams, to run a maximum likelihood analysis of the position of the pulsar. The \texttt{SEEKAT} maximum likelihood position for 47 Tuc ai was found  to be at right ascension $\alpha$ = 0:24:03.37 $_{-0.52} ^{+0.54}$ and declination $\delta$ = $-$72:04:56.00 $_{-1.90} ^{+1.80}$, where the uncertainties are at 1-$\sigma$ confidence level. Fig. \ref{fig:Localization} shows the localized position of the pulsar with respect to the cluster centre.

\subsection{Pulse profile evolution with frequency}\label{sec:prof_evol}

\begin{figure}
    \centering
    
    \includegraphics[width= 1\linewidth]{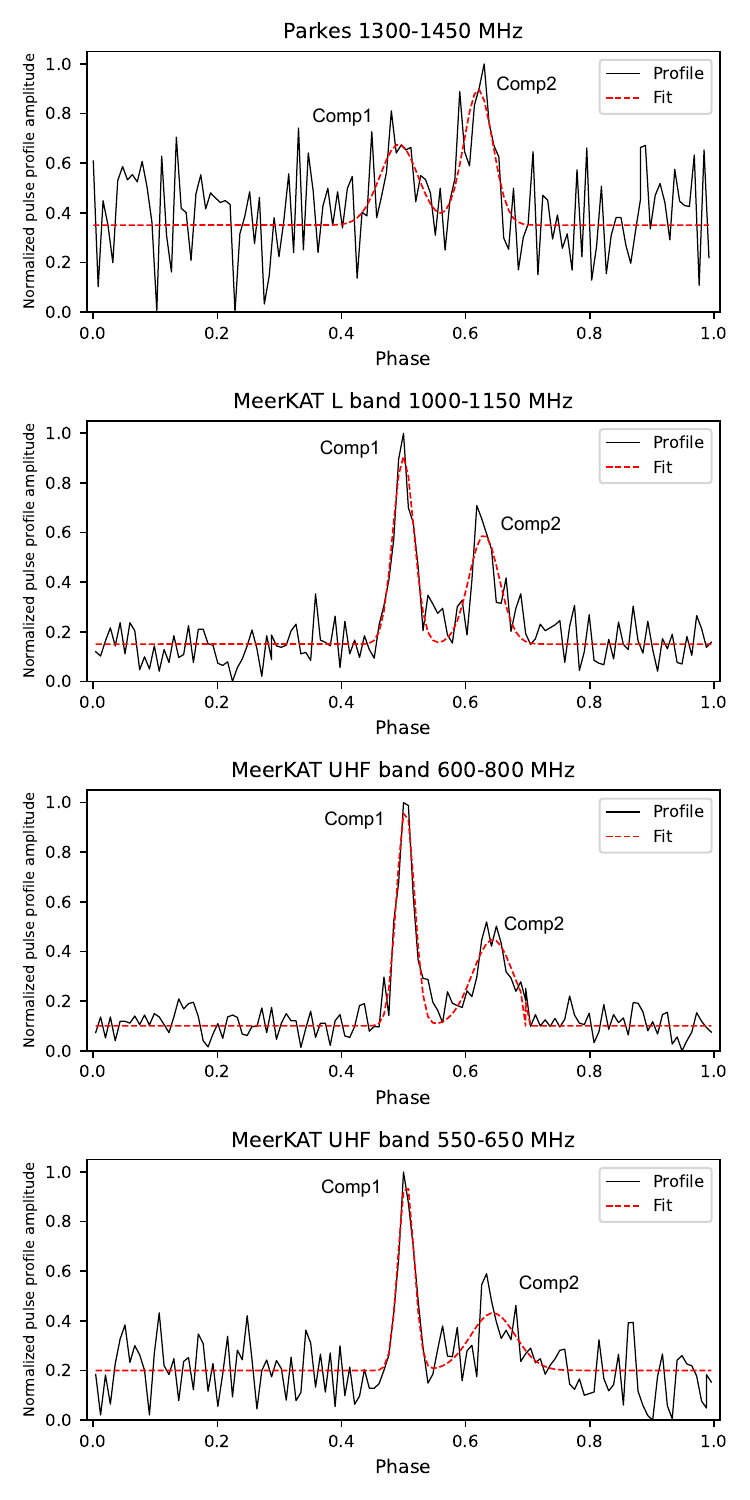}
    \caption{Evolution of pulse profile of PSR 47 Tuc ai with observing frequency. The profile components are marked with Comp1 and Comp2. Each subplot shows the pulse profile at different observing frequencies in black and 2-component Gaussian fits to the profile in dotted red lines. The profiles have been normalized in the amplitude. The results obtained from fit - relative amplitudes, widths and separation between the components are reported in Table \ref{table:prof_evol}. }
    \label{fig:combined_profiles}
\end{figure}

\begin{figure}
    
    \includegraphics[width=1\linewidth, height =1\linewidth]{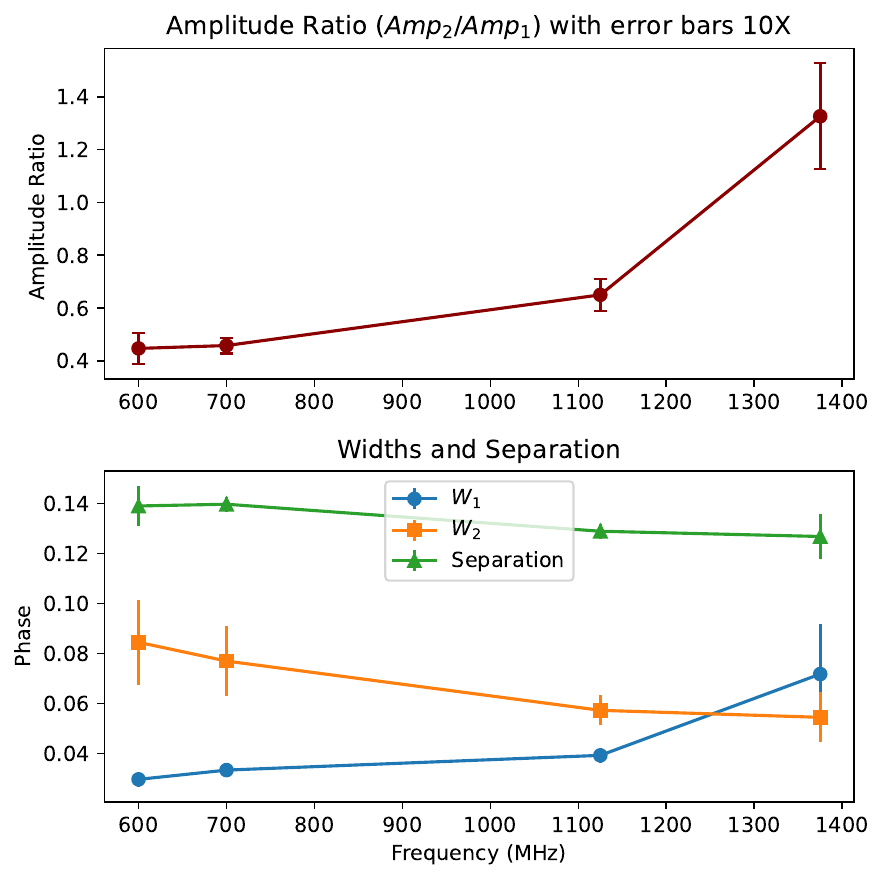}
    \caption{Variation of relative amplitudes, widths and separation between two components of the pulse profile with respect to observing frequency. The fitted profiles at different frequency are shown in Fig. \ref{fig:combined_profiles} ans fitting statistics are listed in table \ref{table:prof_evol}. It can be seen from the upper subplot that the relative amplitude of first component increases as a function of observing frequency.}
    \label{fig:freq_vs_components}
\end{figure}

A strong variation of the integrated pulse profile with observing frequency is observed for 47 Tuc ai. This phenomenon is documented in Figs. \ref{fig:combined_profiles} and \ref{fig:freq_vs_components}, where we show the brightest detection of the pulsar as seen in different sub-bands of the MeerKAT L, MeerKAT UHF and Parkes 1300-MHz bands. In Fig. \ref{fig:combined_profiles}, it can be clearly seen that the relative height between the two components changes as a function of frequency, with the second component becoming brighter with increasing frequency.

We quantified this effect by fitting each of the two components with a gaussian and by measuring their amplitudes, widths and relative separation in each sub-band. The results of the fitting are summarized in Table \ref{table:prof_evol} and shown graphically in Fig. \ref{fig:freq_vs_components}.

Here we do not discuss the implications of this effect, as these are beyond the scope of this paper. 
\begin{comment}

\end{comment}

\begin{table*}[!htbp]
\centering
\caption{Profile evolution of pulsar 47 Tuc ai.   }
\begin{tabular}{lllll}
\hline
Frequency  & $\frac{Amp_2 }{Amp_1 }$ & $W_1$ & $W_2$& Sep  \\
(MHz)  & & & &(phase)\\
\hline
550-650   & 0.45  $\pm$ 0.06 & 0.030 $\pm$ 0.003  & 0.084 $\pm$ 0.017& 0.139  $\pm$ 0.008\\
600-800   & 0.46 $\pm$ 0.03  & 0.033 $\pm$ 0.002  & 0.077 $\pm$ 0.014& 0.140  $\pm$ 0.003\\ 
1100-1150 & 0.65 $\pm$ 0.06  & 0.039 $\pm$ 0.003  & 0.057 $\pm$ 0.006 & 0.129 $\pm$ 0.003\\
1300-1450 & 1.33 $\pm$ 0.20  & 0.072 $\pm$ 0.020  & 0.055 $\pm$ 0.010 & 0.127 $\pm$ 0.009\\
\hline
\end{tabular}
\tablefoot{$\frac{Amp_2 }{Amp_1 }$ represents the ratio relative amplitudes of two components of the profile, $W_1$ and $W_2$ are the FWHMs of gaussian functions fitted to the profile components and the last column gives the separation between the peaks of the profile in terms of phase units. }
\label{table:prof_evol}
\end{table*}

\subsection{Timing}\label{sec:timing}
After the determination and refinement of the orbital parameters, we proceeded with the timing analysis for 47 Tuc ai, attempting to derive the phase-connected timing solution (i.e. an ephemeris that can unambiguously count every rotation of the pulsar over the time-span of the data).

We used the software \texttt{dspsr}\footnote{\url{https://dspsr.sourceforge.net/}}  \citep{2011PASA...28....1vanStraten} to fold the search-mode data of all the observations in which the pulsar was detected. We did this using an ephemeris containing the rotational and orbital parameters, obtained as described in the previous section.

The folded archive files were then cleaned from radio frequency interference (RFI) by using the \texttt{pazi} routine of the  \texttt{PSRCHIVE}\footnote{\url{https://psrchive.sourceforge.net/}} software \citep{2004PASA...21..302Hotan}, which was re-called from \texttt{alex\_clean\_archives}\footnote{\url{https://github.com/alex88ridolfi/PSRALEX/blob/master/alex_clean_archives}}, a python-based wrapper of the program that helps automate the cleaning of large datasets.

We then created a standard profile using the detection of 47 Tuc ai of the MeerKAT UHF observation 26U1, as the pulsar was detected in it with the highest signal-to-noise ratio (S/N). The integrated pulse profile of that detection was fitted with von Mises functions using \texttt{paas} from \texttt{PSRCHIVE} to obtain a noise-free template. 

The topocentric pulse times of arrival (ToAs) were then extracted by cross-correlating the template against an appropriate set of sub-integrations, and possibly sub-bands from each folded archive. This was done using \texttt{alex\_toasel}\footnote{\url{https://github.com/alex88ridolfi/PSRALEX/blob/master/alex\_toasel}}, another python-based wrapper that calls the \texttt{PSRCHIVE}'s \texttt{pat} routine.

\begin{figure*}
    \centering
    \includegraphics[width=0.8\linewidth]{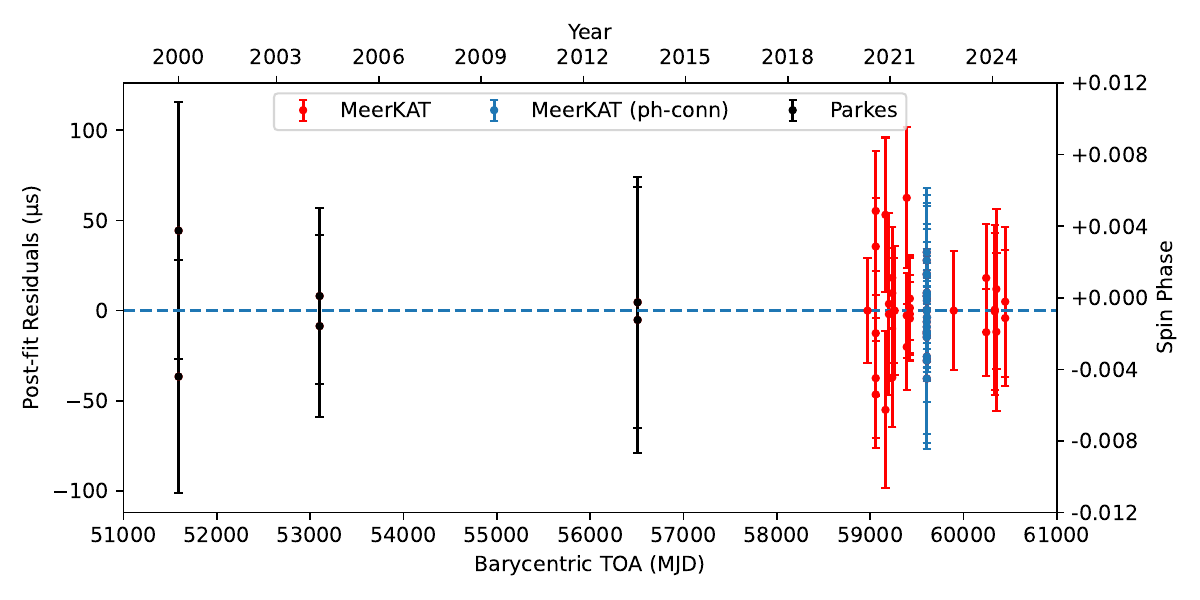}
    \caption{Post-fit timing residuals of 47 Tuc ai as per the timing solution in table \ref{table:timing_sol}. Groups of ToAs are marked in different colours as follows: Red - MeerKAT  (jumped); Blue - MeerKAT phase-connected ; Black - Parkes Multibeam.}
    \label{fig:timing_residuals}
\end{figure*}

The ToAs were fitted for the timing model parameters using the  \texttt{TEMPO}\footnote{\url{https://github.com/nanograv/tempo}} pulsar timing software \citep{2015ascl.soft09002Nice_tempo}. Initially, we kept 'JUMP' statements  between consecutive groups of locally phase-connected ToAs from the individual observations. These JUMPs allow \texttt{TEMPO} to fit for arbitrary time offsets between the groups of ToAs.
MeerKAT observations 25U to 28U (see Table \ref{table:observations}) were part of a 2nd observing campaign made over 3 days. These had frequent bright detections of 47 Tuc ai which were manually phase connected easily. This means that we could start determining the number of rotations {\em between} the observations without any ambiguities. We attempted to do so with \texttt{DRACULA}\footnote{\url{https://github.com/pfreire163/Dracula}}, a software developed by \citet{2018MNRAS.476.4794Freire_dracula}, which automatically determines the rotation counts of pulsars starting from the given input ephemeris. Since there were only 59 ToAs spanning the timing baseline of nearly 25 years, the dataset was sparse with irregular detections and large gaps of tens of years with no ToAs. Due to this limitation of the dataset, we could not find an unambiguous phase-connected timing solution. The results presented here have ToAs from observations 25U to 28U manually connected, and the rest are jumped. The resulting parameters are presented in Table~\ref{table:timing_sol} and the post-fit timing residuals are plotted in Fig. \ref{fig:timing_residuals}. 

Any MeerKAT data taken with APSUSE system in UHF band before 21 Jan 2022 and APSUSE L band taken before 28 Jan 2021 may occasionally suffer positive time offsets due to heap losses in the data recoding. These time offsets cannot be trivially determined. Hence, the ToAs generated from the data in observations 10L, 11L, 12U, 15L, 17L, 18U, 20U and 21U could not be trustworthily included while attempting for the phase-connection, and they were necessarily jumped. In the future, while attempting for the phase connection, it is important to compensate for this issue.

\begin{table}
\caption{Parameters of 47~Tuc~ai from the partially connected timing solution. }
\centering
\begin{footnotesize}

\begin{tabular}{lc}
%\hline \noalign{\vskip 1pt}
\multicolumn{2}{c}{Timing parameters for PSR J0024$-$7204ai}   \\
\hline \noalign{\vskip 1pt}
\multicolumn{2}{c}{Assumed parameters}  \\
\hline \noalign{\vskip 1pt}
Phase connection & Partial \\
Reference epoch (MJD) & 59606.213739  \\
Start of timing data (MJD) & 51591.160 \\
End of timing data (MJD) & 60446.212  \\
Solar System Ephemeris & DE440 \\
Terrestrial time standard & UTC(NIST)   \\
Time Units & TDB  \\
Number of ToAs & 59\\
Binary Model & DD   \\
Residuals r.m.s. ($\upmu$s) &  23.969  \\
Reduced $\chi^2$, $\chi^2_{\rm red} \equiv \chi^2/n_\mathrm{free}$ & 0.82 \\
Right Ascension, $\alpha$ (J2000) &  00:24:03.37  \\
Declination, $\delta$ (J2000) &  $-$72:04:56.0 \\
Proper motion in $\alpha$, $\mu_{\alpha} \cos\delta$ (mas $\rm yr^{-1}$)  & 5.00   \\
Proper motion in $\delta$, $\mu_{\delta}$ (mas $\rm yr^{-1}$)  & $-$2.84 \\[3pt]
\hline \noalign{\vskip 1pt}
\multicolumn{2}{c}{Fitted Parameters} \\
\hline \noalign{\vskip 1pt}
Spin Frequency, $f$ (Hz) & 76.76528761(2)   \\
Dispersion Measure, DM (pc cm$^{-3}$) &24.366(6)   \\
Projected Semi-major Axis, $x_\mathrm{p}$ (s) &  5.35463(1)   \\
Orbital Eccentricity, $e$ & 0.179103(5) \\
Epoch of passage at Periastron, $T_0$ (MJD) &  59163.18181(1)   \\
Orbital Period, $P_{\rm b}$ (days) & 1.65332143(3)  \\
Longitude of Periastron, $\omega$ (deg) &  224.824(3)  \\
Rate of advance of periastron, $\dot\omega$ (deg/yr) & 0.1601(46)  \\[3pt]
\hline \noalign{\vskip 1pt}
\multicolumn{2}{c}{Derived Parameters} \\
\hline \noalign{\vskip 1pt}
Spin period, $P$  (s) & 0.01302672122921(1)  \\

Offset from GC centre, $\theta_\perp$ (arcsec) & 17.03   \\
Offset from GC centre, $\theta_\perp$ (core radii)  & 0.76  \\
Mass function, $f(M_\mathrm{p})$ (M$_\odot$) & 0.0603   \\

Total mass of the system, $M_\mathrm{tot}$, (M$_\odot$) &2.41(11)  \\
Minimum companion mass, $M_{\rm c,min}$, (M$_\odot$) & 0.7   \\
Maximum pulsar mass, (M$_\odot$) & 1.7   \\ [3pt]
\hline
\end{tabular}
\end{footnotesize}
\tablefoot{The proper motions were kept fixed to the cluster mean value from \citet{2017MNRAS.471..857Freire} and position was kept fixed to SEEKAT localization (sec. \ref{sec:localization}).  The numbers in parenthesis are 1-$\sigma$ uncertainties on the respective values. The minimum companion mass and maximum
pulsar mass are derived from the total mass and the mass function. The timing residuals are plotted in Fig. \ref{fig:timing_residuals}.}
\label{table:timing_sol}
\end{table}

\subsection{Measurement of the mass function and $\dot\omega$}\label{sec:Omdot}
Knowing the Keplerian parameters, we can immediately calculate the binary mass function as follows: 
\begin{equation}\label{eq:mass_function}
f(M_{\mathrm{p}})
\equiv
\frac{x_{\mathrm{p}}^{3}}{T_\odot}
\frac{4\pi^{2}}{P_{\mathrm{b}}^{2}}
=
\frac{(M_{\mathrm{c}}\sin i)^{3}}{(M_{\mathrm{p}} + M_{\mathrm{c}})^{2}} = 0.0603... \, \rm M_\odot\, ,
\end{equation}
where $T_{\odot}$ =  $\mathcal{G} \mathcal{M}_{\odot}^N$ /$c^3$ = 4.925490947... ${\upmu}$s is an exact quantity, the {\em solar mass parameter} \citep{2016AJ....152...41Prsa} in time units, and 
$M_p$, $M_c$ are the masses of pulsar and the companion, respectively. In any equation with $T_\odot$, the masses are adimensional, we add the unit $\rm M_{\odot}$ to make it clear that these are multiples of the solar mass parameter.

Despite the lack of a phase-connected timing solution, the combination of the large eccentricity and large semi-major axis of the system allowed a highly significant estimate of the rate of advance of the periastron, $\dot\omega$ = 0.1601 $\pm$ 0.0046 deg yr$^{-1}$. This precise measurement greatly benefited from the addition of Parkes ToAs from the year 2000, 2004 and 2013, making for a total 25-year timing baseline.

\begin{figure*}
    \centering
    \includegraphics[width=0.75\linewidth]{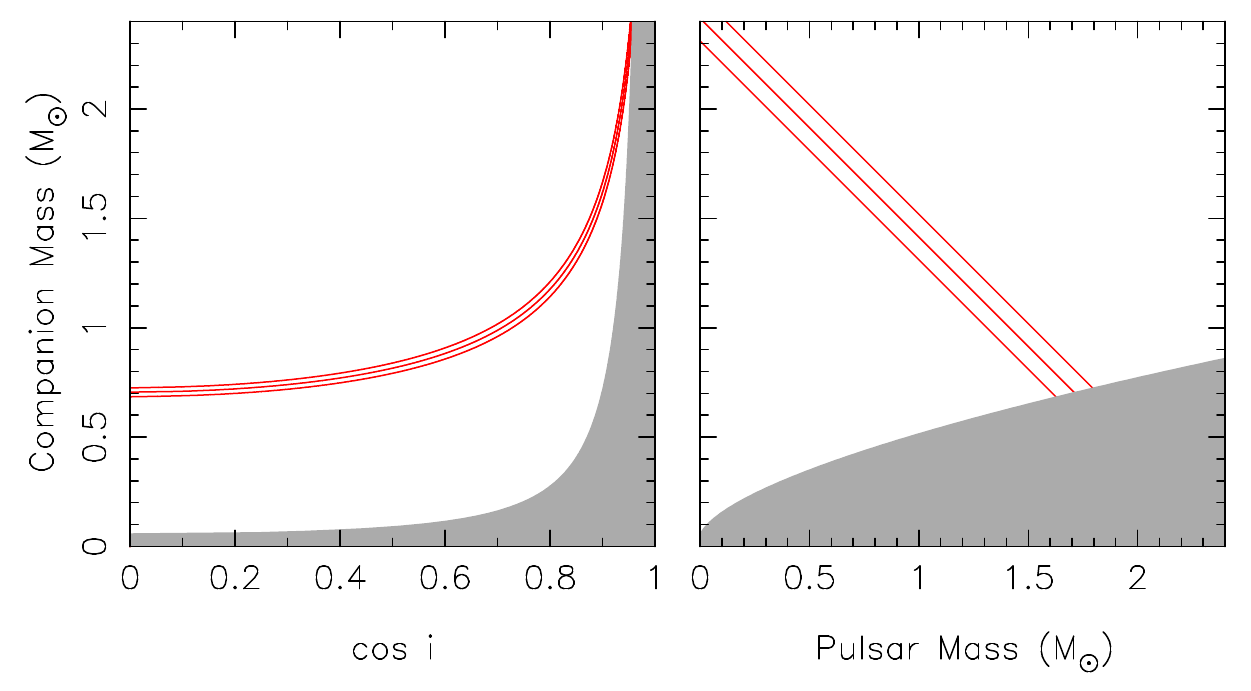}
    \caption{Mass-Mass diagram of 47 Tuc ai. The plot on left shows  $M_\mathrm{c}$ as a function of $\cos i$, while $i$ is inclination of the system and the one on the right shows $M_\mathrm{c}$ as a function of $M_\mathrm{p}$. The gray region in the left plot is excluded, as it implies negative $M_\mathrm{p}$. For the plot on the right side, gray region is excluded by the mass function and the constraint that sin $i$ $\leq$ 1. Red lines in both the plots show nominal values and total mass derived from $\dot\omega = 0.1601 \pm 0.0046 \deg \mathrm{yr}^{-1}$ (see section \ref{sec:Omdot}). From these, we infer that $M_{ \rm tot}$ = 2.41 $\pm 0.11$ M$_{\odot}$,  $M_{\rm c,min}$ $\geq$ 0.7 M$_{\odot}$, and  $M_{\rm p,max}$ $\leq$ 1.7 M$_{\odot}$. }
    \label{fig:mass_mass_diag}
\end{figure*}

Assuming it is a fully relativistic effect, then in general relativity it is given by \citep{Robertson_1938_twobodyprobinGR,1982ApJ...253..908Taylor}:
\begin{equation}
    \dot\omega = 3 \space ( T_{\odot} M_{\rm tot})^{2/3} \space \left(\frac{2\pi}{P_{\rm b}}\right)^{5/3} \space \frac{1}{1 - e^2},
\end{equation}
where $M_{\rm tot}$ is the total mass of the system (pulsar plus companion). The measured value of $\dot \omega$ implies $M_{\rm tot}$ = 2.41 $\pm$ 0.11 $\rm M_{\odot}$. In combination with the mass function, this results in constraints on the pulsar and companion masses (Fig. \ref{fig:mass_mass_diag}): the maximum pulsar mass ($M_{\rm p,max}$) is 1.7 $\rm M_{\odot}$  and the minimum companion mass ($M_{\rm c,min}$) is 0.7 $\rm M_{\odot}$. For $M_{\rm p}$ = 0, the minimum $\sin i$ is 0.284 which implies $i \geq$ 16.5\,\textdegree.

\section{Nature and origin of the system}\label{sec:nature_and_origin}
\begin{figure}
    \includegraphics[width=1\linewidth, height =1\linewidth]{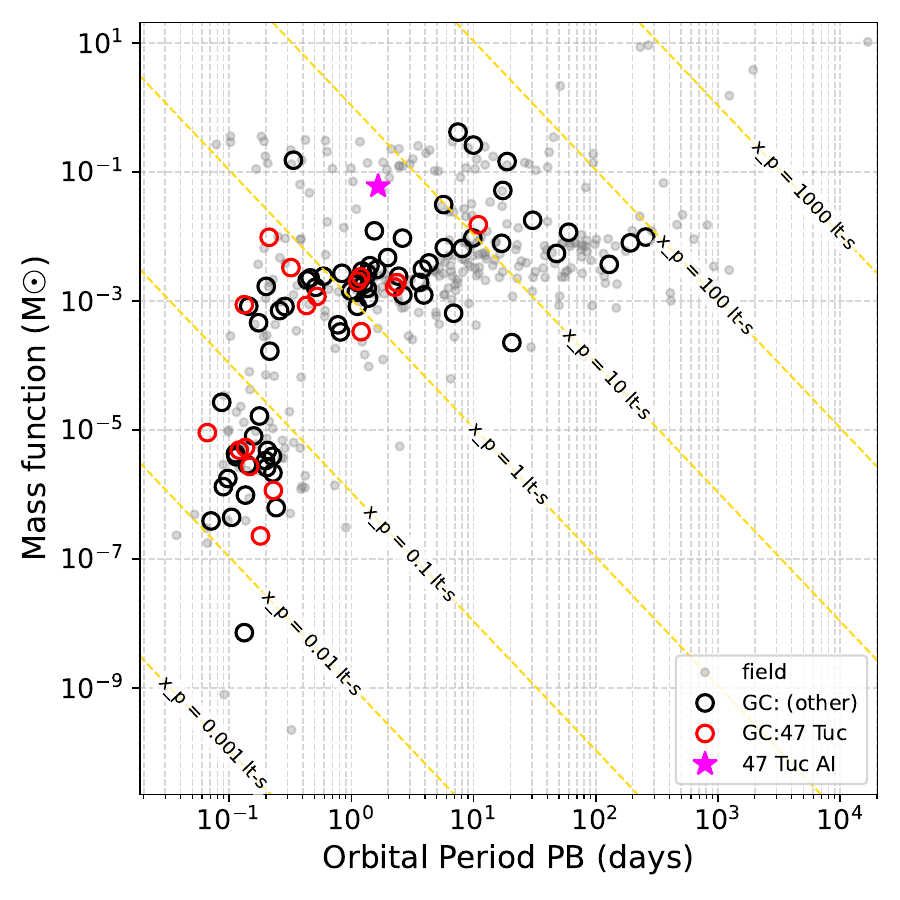}
    \caption{Mass function plotted against orbital period for all the binary pulsars from the ATNF pulsar catalogue\protect\footnotemark \citep{2005AJ....129.1993Manchester}. The binary pulsars in the Galactic field are shown with filled gray dots, while the pulsars in GC are circles. Pulsars in GC 47 Tuc are in red, and the rest in other GCs are shown with black circles. The inclined yellow dotted lines represent constant projected semi-major axis of the binary orbit, of which, smallest on the lower left side is $x$ = 1 lt-millisecond. The successive lines are order of magnitude higher than the previous one. Magenta coloured star symbol represents 47 Tuc ai system.  }
    \label{fig:mass_funcVsPb}
\end{figure}
\footnotetext{\url{https://www.atnf.csiro.au/research/pulsar/psrcat/}} 

The previously known pulsar population in 47 Tuc closely resembled the population of fast-spinning MSPs in the Galactic disk \citep{2016MNRAS.462.2918Ridolfi, 2017MNRAS.471..857Freire}: all pulsars have a low-mass companions with orbits that are circular or with a low eccentricity, although in the case of 47 Tuc H and to a lesser extent 47 Tuc E this is significantly raised by interactions with other stars. 

Pulsar 47 Tuc ai qualifies simultaneously for three records within the pulsar population currently known in 47 Tuc: (a) it is the slowest spinning and only one mildly recycled pulsar; (b) its orbital eccentricity is highest among all the binary pulsars; (c) its companion is most massive among all the binary pulsar companions. It clearly is different from other MSP - Helium WD systems in the cluster, since for this orbital period the mass of a He WD companion should be about 0.2 M$_\odot$ \citep{1999A&A...350..928Tauris}.

The large system mass ($M_{\rm tot} = 2.41 \pm 0.11\,\mathrm{M}_\odot$), companion
mass ($M_\mathrm{c} > 0.7 \, \rm M_{\odot}$) and the orbital eccentricity are close to what one finds among double neutron star \citep[DNS, for a review see][]{2017ApJ...846..170Tauris} systems, as we can see in Fig.~\ref{fig:mass_funcVsPb}, where 47 Tuc ai appears with a magenta colour near a horizontal clump where the Galactic DNSs are located. Recently, one such system has been found in a globular cluster, M71D \citep{2025ApJS..279...51Lian}, which is thought to have formed from the original population of massive stars in that GC.
However, despite the fact that the total mass is still 1-$\sigma$ consistent with the lightest DNSs in the Galaxy \citep{2017ApJ...851L..29Martinez,2025A&A...704A.153Meng}, the nominal value indicates a lower total mass. Furthermore, the  
spin period is shorter than the 17 ms observed for PSR~J1946+2052, the fastest-spinning pulsar in a DNS known \citep{2018ApJ...854L..22Stovall,2025A&A...704A.153Meng}.

Another possibility is that the system is an exchange encounter. Several exchange encounters have been found in GCs that superficially resemble DNSs, like PSR~B2127+11C in the GC M15 ($P = 30.5 \, \mathrm{ms}, P_\mathrm{b} = 8\, \mathrm{h},~e = 0.68$).
When the characteristic age this pulsar was measured, \cite{1991ApJ...374L..41Prince} suggested that the system likely had a low-mass companion at an early stage, which partially recycled it, and that at a later stage the pulsar acquired its current, more massive companion in an exchange interaction. This massive degenerate companion can therefore be either a massive WD or a NS. Thus, dynamically, PSR~B2127+11C is a secondary exchange encounter.
However, this origin for 47 Tuc ai is also unlikely, because systems going through an exchange encounter typically retain higher eccentricities (for systems where it is clearer that this is the origin, $0.38 < e < 0.90$); the orbital eccentricity of PSR~B2127+11C is typical of such systems. 
Furthermore, secondary exchange encounters are most likely to form in GCs with a high $\gamma$, for 47~Tuc this number is much lower than those of M15, NGC~6544 or NGC~1851, where undoubted secondary exchange products have been found.

Therefore, the most likely possibility is that 47~Tuc~ai a mildly recycled pulsar with a massive WD companion. The progenitors of massive companions like CO or ONeMg WDs have shorter evolutionary phases, which result in spin periods like those observed in 47 Tuc ai. The orbits in this case are circularized with eccentricities of the order of 10$^{-2}$ to 10$^{-5}$, which is much lower than the observed eccentricity.
However, this is not a problem: in the cluster cor
stellar encounters in the cluster core are very frequent. The gravitational perturbations by multiple close stellar dynamical encounters can induce the observed orbital eccentricity in a binary system. Assuming that the birth eccentricity was zero, an estimate of the time required for the cluster to produce a system this eccentric can be made from eq. given by \citet{1995ApJ...445L.133Rasio},
\begin{equation}\label{eq:t_greater_e_eq}
\begin{split}
    t_{>e} \simeq & \; 2 \times 10^{11} \ \mathrm{yr} \ 
    \left( \frac{n}{10^4 \ \mathrm{pc}^{-3}} \right)^{-1} 
    \left( \frac{v_0}{10 \ \mathrm{km \ s^{-1}}} \right) \\
    & \times \left( \frac{P_{\rm b}}{\mathrm{days}} \right)^{-2/3} 
    \left[-\ln\ (e/4)\right]^{-2/3},
\end{split}
\end{equation}
where the equation is valid for $e \geq 0.01$. In this eq.,  $n$ is number density, $v_0$ is one-dimensional velocity dispersion. To estimate $n$, we used the stellar number density profile of 47 Tuc given by \citet{2013ApJ...774..151Miocchi}. The stellar number density per arcsecond at $\approx$ 17\arcsec (the angular offset of pulsar from the cluster center) at the distance of 4.69 kpc distance, translates to $n$ = 4.0 $\times 10^5$ $ \rm{pc}^{-3}$ . We adopted $v_0 = 11.0 \, \mathrm{km\, s}^{-1}$  from the catalogue of \citet{1996AJ....112.1487Harris} (version 2010, \citealt{2010arXiv1012.3224Harris}). This yields $t_{>e}$ $\approx$ 1.85 GYr, which is well within the range of characteristic ages estimated for other pulsars in this GC \citep{2017MNRAS.471..857Freire} and shows clearly that this system could have acquired is orbital eccentricity via this mechanism well within the age of the globular cluster.

As noted in the introduction, the massive progenitors to these massive WDs left the main sequence in the initial few Gyr of the age of the GC. This means that it is likely that the system is very old, with a characteristic age similar or larger than the age of the globular cluster.

\section{Conclusion and prospects}\label{sec:conclusion}

In this paper, we presented the  timing analysis of 47 Tuc ai, a binary pulsar system recently discovered by the MeerKAT radio telescope.  
This system stands out from the rest of the pulsar population in 47 Tuc including the recent MeerKAT discoveries, being the first and only known mildly recycled pulsar ($P$ $\approx$ 13.026 ms) in the cluster. The orbital eccentricity of $\approx$ 0.18 of this system is highest among the 47 Tuc binary pulsars. 
Due to a strong time and frequency dependent scintillation in the direction of the cluster, the evolution of folded profile of pulsar with respect to the observing frequency was uncovered and quantified using the brightest detections in MeerKAT L, UHF and Parkes archival observations.    

For the pulsar timing analysis, we also included archival data from the Parkes radio telescope, giving us the oldest detection of this pulsar in the year 2000. Although the data spans nearly 25 years, the time- and frequency-dependent scintillation in the direction of 47 Tuc made it challenging to have high S/N ratio detections of the pulsar, yielding 59 total ToAs only. The timing solution is partially connected. 

One of the orbital parameters is the rate of advance of periastron, $\dot \omega = 0.1601 \pm 0.0046~ \rm deg~yr^{-1}$. The total system mass, assuming the effect is entirely due to general relativity, is $2.41 \pm 0.11 \mathrm{M}_\odot$. Combined with the pulsar mass function, this implies $M_\mathrm{p} \lesssim 1.7~ \mathrm{M}_\odot$ and $M_\mathrm{c} \gtrsim 0.7~ \mathrm{M}_\odot$.

Continued timing will almost certainly result in the determination of a phase-connected timing solution. This will allow further characterisation of the system: the precise localization might allow the detection of the massive companion; precise estimates of the spin frequency derivative and orbital period derivative will allow a precise estimate of the intrinsic spin-down of the pulsar (and thus magnetic field and characteristic age), which will be important to test the nature of the system. In particular, the main hypothesis we made above on the nature of the system predicts that the characteristic age should not be smaller than the eccentricity timescale $t_{>e}$, otherwise the eccentricity of the system has not originated in close encounters with other stars in the cluster,  but instead comparable or larger than the age of the GC, since the massive progenitors to these
massive WDs were only present at the early stages of the cluster's evolution. The characteristic age will eventually be measurable when the variation of the orbital period is measured with sufficient precision, as for many of the binaries in 47~Tuc \citep{2017MNRAS.471..857Freire}.

No additional relativistic effects are currently measurable; however, there are good prospects for the detection of the Einstein delay and the determination of the individual masses in the future, which will also be important for an improved characterisation of this system. 

\begin{acknowledgements}
The MeerKAT telescope is operated by the South African Radio Astronomy Observatory, which is a facility of the National Research Foundation, an agency of the Department of Science and Innovation. SARAO acknowledges the ongoing advice and calibration of GPS systems by the National Metrology Institute of South Africa (NMISA) and the time space reference systems department of the Paris Observatory. The TRAPUM observations used the FBFUSE and APSUSE computing clusters for data acquisition, storage and analysis. These clusters were funded and installed by the Max-Planck-Institut für Radioastronomie and the Max-Planck-Gesellschaft. MeerTime data is housed on the OzSTAR supercomputer at Swinburne University of Technology. PTUSE was developed with support from the Australian SKA Office and Swinburne University of
Technology. This work has made use of CSIRO Data Access Portal (https://data.csiro.au/domain/atnf) and authors acknowledge CSIRO and ATNF for keeping up this facility.
DR, AR, PCCF, WC, PVP acknowledge continuing valuable support from the Max-Planck Society. DR  and PCCF also acknowledge INAF Cagliari and the organizing committee of 'Pulsar 2025 - A conference in memory of Nichi D'Amico', that encouraged discussions improving the quality of this work. FA acknowledges that part of the research activities described in this paper were carried out with the contribution of the NextGenerationEU funds within the National Recovery and Resilience Plan (PNRR), Mission 4 – Education and Research, Component 2 – From Research to Business (M4C2), Investment Line 3.1 – Strengthening and creation of Research Infrastructures, Project IR0000034 – ‘STILES -Strengthening the Italian Leadership in ELT and SKA’. Part of this work has been funded using resources from the INAF Large Grant 2022 "GCjewels" (P.I. Andrea Possenti) approved with the Presidential Decree 30/2022. AP contribution was also supported  by the "Italian Ministry of Foreign Affairs and International Cooperation", grant number ZA23GR03, under the project "RADIOMAP- Science and technology pathways to MeerKAT+: the Italian and South African synergy".

\end{acknowledgements}

\bibliography{references}

@INPROCEEDINGS{Stappers_Kramer2016,
       author = {{Stappers}, B. and {Kramer}, M.},
        title = "{An Update on TRAPUM}",
    booktitle = {MeerKAT Science: On the Pathway to the SKA},
         year = 2016,
        month = jan,
          eid = {9},
        pages = {9},
          doi = {10.22323/1.277.0009},
       adsurl = {https://ui.adsabs.harvard.edu/abs/2016mks..confE...9S}
}

@ARTICLE{2023RASTI...2..114Bezuidenhout,
       author = {{Bezuidenhout}, M.~C. and {Clark}, C.~J. and {Breton}, R.~P. and {Stappers}, B.~W. and {Barr}, E.~D. and {Caleb}, M. and {Chen}, W. and {Jankowski}, F. and {Kramer}, M. and {Rajwade}, K. and {Surnis}, M.},
        title = "{Tied-array beam localization of radio transients and pulsars}",
      journal = {RAS Techniques and Instruments},
         year = 2023,
        month = jan,
       volume = {2},
       number = {1},
        pages = {114-128},
          doi = {10.1093/rasti/rzad007},
archivePrefix = {arXiv},
       eprint = {2302.09812},
 primaryClass = {astro-ph.HE},
       adsurl = {https://ui.adsabs.harvard.edu/abs/2023RASTI...2..114B}
}

@ARTICLE{2025A&A...697A.166Dutta,
       author = {{Dutta}, A. and {Freire}, P.~C.~C. and {Gautam}, T. and {Wex}, N. and {Ridolfi}, A. and {Champion}, D.~J. and {Venkatraman Krishnan}, V. and {Rosie Chen}, C. -H. and {Cadelano}, M. and {Kramer}, M. and {Abbate}, F. and {Bailes}, M. and {Balakrishnan}, V. and {Corongiu}, A. and {Gupta}, Y. and {Padmanabh}, P.~V. and {Possenti}, A. and {Ransom}, S.~M. and {Zhang}, L.},
        title = "{NGC 1851A: Revealing an ongoing three-body encounter in a dense globular cluster}",
      journal = {\aap},
         year = 2025,
        month = may,
       volume = {697},
          eid = {A166},
        pages = {A166},
          doi = {10.1051/0004-6361/202452433},
archivePrefix = {arXiv},
       eprint = {2503.05466},
 primaryClass = {astro-ph.HE},
       adsurl = {https://ui.adsabs.harvard.edu/abs/2025A&A...697A.166D}
}

@ARTICLE{2024Sci...383..275Barr,
       author = {{Barr}, Ewan D. and {Dutta}, Arunima and {Freire}, Paulo C.~C. and {Cadelano}, Mario and {Gautam}, Tasha and {Kramer}, Michael and {Pallanca}, Cristina and {Ransom}, Scott M. and {Ridolfi}, Alessandro and {Stappers}, Benjamin W. and {Tauris}, Thomas M. and {Venkatraman Krishnan}, Vivek and {Wex}, Norbert and {Bailes}, Matthew and {Behrend}, Jan and {Buchner}, Sarah and {Burgay}, Marta and {Chen}, Weiwei and {Champion}, David J. and {Chen}, C. -H. Rosie and {Corongiu}, Alessandro and {Geyer}, Marisa and {Men}, Y.~P. and {Padmanabh}, Prajwal Voraganti and {Possenti}, Andrea},
        title = "{A pulsar in a binary with a compact object in the mass gap between neutron stars and black holes}",
      journal = {Science},
         year = 2024,
        month = jan,
       volume = {383},
       number = {6680},
        pages = {275-279},
          doi = {10.1126/science.adg3005},
archivePrefix = {arXiv},
       eprint = {2401.09872},
 primaryClass = {astro-ph.HE},
       adsurl = {https://ui.adsabs.harvard.edu/abs/2024Sci...383..275B}
}

@BOOK{2023pbse.book.....Tauris,
       author = {{Tauris}, Thomas M. and {van den Heuvel}, Edward P.~J.},
        title = "{Physics of Binary Star Evolution. From Stars to X-ray Binaries and Gravitational Wave Sources}",
         year = 2023,
          doi = {10.48550/arXiv.2305.09388},
       adsurl = {https://ui.adsabs.harvard.edu/abs/2023pbse.book.....T}
}

@ARTICLE{1999A&A...350..928Tauris,
       author = {{Tauris}, Thomas M. and {Savonije}, Gerrit J.},
        title = "{Formation of millisecond pulsars. I. Evolution of low-mass X-ray binaries with P\_orb> 2 days}",
      journal = {\aap},
         year = 1999,
        month = oct,
       volume = {350},
        pages = {928-944},
          doi = {10.48550/arXiv.astro-ph/9909147},
archivePrefix = {arXiv},
       eprint = {astro-ph/9909147},
 primaryClass = {astro-ph},
       adsurl = {https://ui.adsabs.harvard.edu/abs/1999A&A...350..928T}
}

@ARTICLE{2004ApJ...606L..53Freire,
       author = {{Freire}, Paulo C. and {Gupta}, Yashwant and {Ransom}, Scott M. and {Ishwara-Chandra}, C.~H.},
        title = "{Giant Metrewave Radio Telescope Discovery of a Millisecond Pulsar in a Very Eccentric Binary System}",
      journal = {\apjl},
         year = 2004,
        month = may,
       volume = {606},
       number = {1},
        pages = {L53-L56},
          doi = {10.1086/421085},
archivePrefix = {arXiv},
       eprint = {astro-ph/0403453},
 primaryClass = {astro-ph},
       adsurl = {https://ui.adsabs.harvard.edu/abs/2004ApJ...606L..53F}
}

@ARTICLE{1964PhRvL..13..789Shapiro,
       author = {{Shapiro}, Irwin I.},
        title = "{Fourth Test of General Relativity}",
      journal = {\prl},
         year = 1964,
        month = dec,
       volume = {13},
       number = {26},
        pages = {789-791},
          doi = {10.1103/PhysRevLett.13.789},
       adsurl = {https://ui.adsabs.harvard.edu/abs/1964PhRvL..13..789S}
}

@ARTICLE{2006ApJ...644L.113Jacoby,
       author = {{Jacoby}, B.~A. and {Cameron}, P.~B. and {Jenet}, F.~A. and {Anderson}, S.~B. and {Murty}, R.~N. and {Kulkarni}, S.~R.},
        title = "{Measurement of Orbital Decay in the Double Neutron Star Binary PSR B2127+11C}",
      journal = {\apjl},
         year = 2006,
        month = jun,
       volume = {644},
       number = {2},
        pages = {L113-L116},
          doi = {10.1086/505742},
archivePrefix = {arXiv},
       eprint = {astro-ph/0605375},
 primaryClass = {astro-ph},
       adsurl = {https://ui.adsabs.harvard.edu/abs/2006ApJ...644L.113J}
}

@ARTICLE{2023ApJ...942L..35B,
       author = {{Balakrishnan}, Vishnu and {Freire}, Paulo C.~C. and {Ransom}, S.~M. and {Ridolfi}, Alessandro and {Barr}, E.~D. and {Chen}, W. and {Krishnan}, Vivek Venkatraman and {Champion}, D. and {Kramer}, M. and {Gautam}, T. and {Padmanabh}, Prajwal V. and {Men}, Yunpeng and {Abbate}, F. and {Stappers}, B.~W. and {Stairs}, I. and {Keane}, E. and {Possenti}, A.},
        title = "{Missing for 20 yr: MeerKAT Redetects the Elusive Binary Pulsar M30B}",
      journal = {\apjl},
         year = 2023,
        month = jan,
       volume = {942},
       number = {2},
          eid = {L35},
        pages = {L35},
          doi = {10.3847/2041-8213/acae99},
archivePrefix = {arXiv},
       eprint = {2301.04983},
 primaryClass = {astro-ph.HE},
       adsurl = {https://ui.adsabs.harvard.edu/abs/2023ApJ...942L..35B}
}

@ARTICLE{1991ApJ...374L..41Prince,
       author = {{Prince}, T.~A. and {Anderson}, S.~B. and {Kulkarni}, S.~R. and {Wolszczan}, A.},
        title = "{Timing Observations of the 8 Hour Binary Pulsar 2127+11C in the Globular Cluster M15}",
      journal = {\apjl},
         year = 1991,
        month = jun,
       volume = {374},
        pages = {L41},
          doi = {10.1086/186067},
       adsurl = {https://ui.adsabs.harvard.edu/abs/1991ApJ...374L..41P}
}

@ARTICLE{2016MNRAS.462.2918Ridolfi,
       author = {{Ridolfi}, A. and {Freire}, P.~C.~C. and {Torne}, P. and {Heinke}, C.~O. and {van den Berg}, M. and {Jordan}, C. and {Kramer}, M. and {Bassa}, C.~G. and {Sarkissian}, J. and {D'Amico}, N. and {Lorimer}, D. and {Camilo}, F. and {Manchester}, R.~N. and {Lyne}, A.},
        title = "{Long-term observations of the pulsars in 47 Tucanae - I. A study of four elusive binary systems}",
      journal = {\mnras},
         year = 2016,
        month = nov,
       volume = {462},
       number = {3},
        pages = {2918-2933},
          doi = {10.1093/mnras/stw1850},
archivePrefix = {arXiv},
       eprint = {1607.07248},
 primaryClass = {astro-ph.HE},
       adsurl = {https://ui.adsabs.harvard.edu/abs/2016MNRAS.462.2918R}
}

@ARTICLE{2001MNRAS.322..885Freire2001a,
       author = {{Freire}, P.~C. and {Kramer}, M. and {Lyne}, A.~G.},
        title = "{Determination of the orbital parameters of binary pulsars}",
      journal = {\mnras},
         year = 2001,
        month = apr,
       volume = {322},
       number = {4},
        pages = {885-890},
          doi = {10.1046/j.1365-8711.2001.04200.x},
archivePrefix = {arXiv},
       eprint = {astro-ph/0010463},
 primaryClass = {astro-ph},
       adsurl = {https://ui.adsabs.harvard.edu/abs/2001MNRAS.322..885F}
}

@ARTICLE{2009MNRAS.395.1775Freire_erratum,
       author = {{Freire}, P.~C. and {Kramer}, M. and {Lyne}, A.~G.},
        title = "{Erratum: Determination of the orbital parameters of binary pulsars}",
      journal = {\mnras},
         year = 2009,
        month = may,
       volume = {395},
       number = {3},
        pages = {1775-1775},
          doi = {10.1111/j.1365-2966.2009.14609.x},
       adsurl = {https://ui.adsabs.harvard.edu/abs/2009MNRAS.395.1775F}
}

@ARTICLE{1996AJ....112.1487Harris,
       author = {{Harris}, William E.},
        title = "{A Catalog of Parameters for Globular Clusters in the Milky Way}",
      journal = {\aj},
         year = 1996,
        month = oct,
       volume = {112},
        pages = {1487},
          doi = {10.1086/118116},
       adsurl = {https://ui.adsabs.harvard.edu/abs/1996AJ....112.1487H}
}

@ARTICLE{2010arXiv1012.3224Harris,
       author = {{Harris}, William E.},
        title = "{A New Catalog of Globular Clusters in the Milky Way}",
      journal = {arXiv e-prints},
         year = 2010,
        month = dec,
          eid = {arXiv:1012.3224},
        pages = {arXiv:1012.3224},
          doi = {10.48550/arXiv.1012.3224},
archivePrefix = {arXiv},
       eprint = {1012.3224},
 primaryClass = {astro-ph.GA},
       adsurl = {https://ui.adsabs.harvard.edu/abs/2010arXiv1012.3224H}
}

@ARTICLE{1990Natur.345..598Manchester,
       author = {{Manchester}, R.~N. and {Lyne}, A.~G. and {D'Amico}, N. and {Johnston}, S. and {Lim}, J. and {Kniffen}, D.~A.},
        title = "{A 5.75-millisecond pulsar in the globular cluster 47 Tucanae}",
      journal = {\nat},
         year = 1990,
        month = jun,
       volume = {345},
       number = {6276},
        pages = {598-600},
          doi = {10.1038/345598a0},
       adsurl = {https://ui.adsabs.harvard.edu/abs/1990Natur.345..598M}
}

@ARTICLE{1995MNRAS.274..547Robinson,
       author = {{Robinson}, Clive and {Lyne}, A.~G. and {Manchester}, R.~N. and {Bailes}, M. and {D'Amico}, N. and {Johnston}, S.},
        title = "{Millisecond pulsars in the globular cluster 47 Tucanae}",
      journal = {\mnras},
         year = 1995,
        month = may,
       volume = {274},
       number = {2},
        pages = {547-554},
          doi = {10.1093/mnras/274.2.547},
       adsurl = {https://ui.adsabs.harvard.edu/abs/1995MNRAS.274..547R}
}

@ARTICLE{2000ApJ...535..975Camilo,
       author = {{Camilo}, F. and {Lorimer}, D.~R. and {Freire}, P. and {Lyne}, A.~G. and {Manchester}, R.~N.},
        title = "{Observations of 20 Millisecond Pulsars in 47 Tucanae at 20 Centimeters}",
      journal = {\apj},
         year = 2000,
        month = jun,
       volume = {535},
       number = {2},
        pages = {975-990},
          doi = {10.1086/308859},
archivePrefix = {arXiv},
       eprint = {astro-ph/9911234},
 primaryClass = {astro-ph},
       adsurl = {https://ui.adsabs.harvard.edu/abs/2000ApJ...535..975C}
}

@ARTICLE{2021MNRAS.504.1407Ridolfi,
       author = {{Ridolfi}, A. and {Gautam}, T. and {Freire}, P.~C.~C. and {Ransom}, S.~M. and {Buchner}, S.~J. and {Possenti}, A. and {Venkatraman Krishnan}, V. and {Bailes}, M. and {Kramer}, M. and {Stappers}, B.~W. and {Abbate}, F. and {Barr}, E.~D. and {Burgay}, M. and {Camilo}, F. and {Corongiu}, A. and {Jameson}, A. and {Padmanabh}, P.~V. and {Vleeschower}, L. and {Champion}, D.~J. and {Chen}, W. and {Geyer}, M. and {Karastergiou}, A. and {Karuppusamy}, R. and {Parthasarathy}, A. and {Reardon}, D.~J. and {Serylak}, M. and {Shannon}, R.~M. and {Spiewak}, R.},
        title = "{Eight new millisecond pulsars from the first MeerKAT globular cluster census}",
      journal = {\mnras},
         year = 2021,
        month = jun,
       volume = {504},
       number = {1},
        pages = {1407-1426},
          doi = {10.1093/mnras/stab790},
archivePrefix = {arXiv},
       eprint = {2103.04800},
 primaryClass = {astro-ph.HE},
       adsurl = {https://ui.adsabs.harvard.edu/abs/2021MNRAS.504.1407R}
}

@ARTICLE{2001MNRAS.326..901Freire_b,
       author = {{Freire}, P.~C. and {Camilo}, F. and {Lorimer}, D.~R. and {Lyne}, A.~G. and {Manchester}, R.~N. and {D'Amico}, N.},
        title = "{Timing the millisecond pulsars in 47 Tucanae}",
      journal = {\mnras},
         year = 2001,
        month = sep,
       volume = {326},
       number = {3},
        pages = {901-915},
          doi = {10.1046/j.1365-8711.2001.04493.x},
archivePrefix = {arXiv},
       eprint = {astro-ph/0103372},
 primaryClass = {astro-ph},
       adsurl = {https://ui.adsabs.harvard.edu/abs/2001MNRAS.326..901F}
}

@ARTICLE{2017MNRAS.471..857Freire,
       author = {{Freire}, P.~C.~C. and {Ridolfi}, A. and {Kramer}, M. and {Jordan}, C. and {Manchester}, R.~N. and {Torne}, P. and {Sarkissian}, J. and {Heinke}, C.~O. and {D'Amico}, N. and {Camilo}, F. and {Lorimer}, D.~R. and {Lyne}, A.~G.},
        title = "{Long-term observations of the pulsars in 47 Tucanae - II. Proper motions, accelerations and jerks}",
      journal = {\mnras},
         year = 2017,
        month = oct,
       volume = {471},
       number = {1},
        pages = {857-876},
          doi = {10.1093/mnras/stx1533},
archivePrefix = {arXiv},
       eprint = {1706.04908},
 primaryClass = {astro-ph.HE},
       adsurl = {https://ui.adsabs.harvard.edu/abs/2017MNRAS.471..857F}
}

@ARTICLE{2001ApJ...557L.105Freire_2001c,
       author = {{Freire}, P.~C. and {Kramer}, M. and {Lyne}, A.~G. and {Camilo}, F. and {Manchester}, R.~N. and {D'Amico}, N.},
        title = "{Detection of Ionized Gas in the Globular Cluster 47 Tucanae}",
      journal = {\apjl},
         year = 2001,
        month = aug,
       volume = {557},
       number = {2},
        pages = {L105-L108},
          doi = {10.1086/323248},
archivePrefix = {arXiv},
       eprint = {astro-ph/0107206},
 primaryClass = {astro-ph},
       adsurl = {https://ui.adsabs.harvard.edu/abs/2001ApJ...557L.105F}
}

@ARTICLE{2005ApJ...630.1029Bogdanov,
       author = {{Bogdanov}, Slavko and {Grindlay}, Jonathan E. and {van den Berg}, Maureen},
        title = "{An X-Ray Variable Millisecond Pulsar in the Globular Cluster 47 Tucanae: Closing the Link to Low-Mass X-Ray Binaries}",
      journal = {\apj},
         year = 2005,
        month = sep,
       volume = {630},
       number = {2},
        pages = {1029-1036},
          doi = {10.1086/432249},
archivePrefix = {arXiv},
       eprint = {astro-ph/0506031},
 primaryClass = {astro-ph},
       adsurl = {https://ui.adsabs.harvard.edu/abs/2005ApJ...630.1029B}
}

@ARTICLE{2021MNRAS.500.1139Hebbar,
       author = {{Hebbar}, P.~R. and {Heinke}, C.~O. and {Kandel}, D. and {Romani}, R.~W. and {Freire}, P.~C.~C.},
        title = "{On the vanishing orbital X-ray variability of the eclipsing binary millisecond pulsar 47 Tuc W}",
      journal = {\mnras},
         year = 2021,
        month = jan,
       volume = {500},
       number = {1},
        pages = {1139-1150},
          doi = {10.1093/mnras/staa3072},
archivePrefix = {arXiv},
       eprint = {2009.13561},
 primaryClass = {astro-ph.HE},
       adsurl = {https://ui.adsabs.harvard.edu/abs/2021MNRAS.500.1139H}
}

@ARTICLE{2017MNRAS.472.3706Bhattacharya,
       author = {{Bhattacharya}, Souradeep and {Heinke}, Craig O. and {Chugunov}, Andrey I. and {Freire}, Paulo C.~C. and {Ridolfi}, Alessandro and {Bogdanov}, Slavko},
        title = "{Chandra studies of the globular cluster 47 Tucanae: A deeper X-ray source catalogue, five new X-ray counterparts to millisecond radio pulsars, and new constraints to r-mode instability window}",
      journal = {\mnras},
         year = 2017,
        month = dec,
       volume = {472},
       number = {3},
        pages = {3706-3721},
          doi = {10.1093/mnras/stx2241},
archivePrefix = {arXiv},
       eprint = {1709.01807},
 primaryClass = {astro-ph.HE},
       adsurl = {https://ui.adsabs.harvard.edu/abs/2017MNRAS.472.3706B}
}

@ARTICLE{2015MNRAS.453.Rivera,
       author = {{Rivera-Sandoval}, L.~E. and {van den Berg}, M. and {Heinke}, C.~O. and {Cohn}, H.~N. and {Lugger}, P.~M. and {Freire}, P. and {Anderson}, J. and {Serenelli}, A.~M. and {Althaus}, L.~G. and {Cool}, A.~M. and {Grindlay}, J.~E. and {Edmonds}, P.~D. and {Wijnands}, R. and {Ivanova}, N.},
        title = "{Discovery of near-ultraviolet counterparts to millisecond pulsars in the globular cluster 47 Tucanae}",
      journal = {\mnras},
         year = 2015,
        month = nov,
       volume = {453},
       number = {3},
        pages = {2707-2717},
          doi = {10.1093/mnras/stv1810},
archivePrefix = {arXiv},
       eprint = {1508.05291},
 primaryClass = {astro-ph.SR},
       adsurl = {https://ui.adsabs.harvard.edu/abs/2015MNRAS.453.2707R}
}

@ARTICLE{2001ApJ...557L..57Edmonds,
       author = {{Edmonds}, Peter D. and {Gilliland}, Ronald L. and {Heinke}, Craig O. and {Grindlay}, Jonathan E. and {Camilo}, Fernando},
        title = "{Optical Detection of a Variable Millisecond Pulsar Companion in 47 Tucanae}",
      journal = {\apjl},
         year = 2001,
        month = aug,
       volume = {557},
       number = {1},
        pages = {L57-L60},
          doi = {10.1086/323122},
archivePrefix = {arXiv},
       eprint = {astro-ph/0107096},
 primaryClass = {astro-ph},
       adsurl = {https://ui.adsabs.harvard.edu/abs/2001ApJ...557L..57E}
}

@ARTICLE{2002ApJ...579..741Edmonds,
       author = {{Edmonds}, Peter D. and {Gilliland}, Ronald L. and {Camilo}, Fernando and {Heinke}, Craig O. and {Grindlay}, Jonathan E.},
        title = "{A Millisecond Pulsar Optical Counterpart with Large-Amplitude Variability in the Globular Cluster 47 Tucanae}",
      journal = {\apj},
         year = 2002,
        month = nov,
       volume = {579},
       number = {2},
        pages = {741-751},
          doi = {10.1086/342985},
archivePrefix = {arXiv},
       eprint = {astro-ph/0207426},
 primaryClass = {astro-ph},
       adsurl = {https://ui.adsabs.harvard.edu/abs/2002ApJ...579..741E}
}

@ARTICLE{2015ApJ...812...63Cadelano,
       author = {{Cadelano}, M. and {Pallanca}, C. and {Ferraro}, F.~R. and {Salaris}, M. and {Dalessandro}, E. and {Lanzoni}, B. and {Freire}, P.~C.~C.},
        title = "{Optical Identification of He White Dwarfs Orbiting Four Millisecond Pulsars in the Globular Cluster 47 Tucanae}",
      journal = {\apj},
         year = 2015,
        month = oct,
       volume = {812},
       number = {1},
          eid = {63},
        pages = {63},
          doi = {10.1088/0004-637X/812/1/63},
archivePrefix = {arXiv},
       eprint = {1509.01397},
 primaryClass = {astro-ph.SR},
       adsurl = {https://ui.adsabs.harvard.edu/abs/2015ApJ...812...63C}
}

@ARTICLE{2002ApJ...581..470Grindlay,
       author = {{Grindlay}, J.~E. and {Camilo}, F. and {Heinke}, C.~O. and {Edmonds}, P.~D. and {Cohn}, H. and {Lugger}, P.},
        title = "{Chandra Study of a Complete Sample of Millisecond Pulsars in 47 Tucanae and NGC 6397}",
      journal = {\apj},
         year = 2002,
        month = dec,
       volume = {581},
       number = {1},
        pages = {470-484},
          doi = {10.1086/344150},
archivePrefix = {arXiv},
       eprint = {astro-ph/0208280},
 primaryClass = {astro-ph},
       adsurl = {https://ui.adsabs.harvard.edu/abs/2002ApJ...581..470G}
}

@ARTICLE{2006ApJ...646.1104Bogdanov,
       author = {{Bogdanov}, Slavko and {Grindlay}, Jonathan E. and {Heinke}, Craig O. and {Camilo}, Fernando and {Freire}, Paulo C.~C. and {Becker}, Werner},
        title = "{Chandra X-Ray Observations of 19 Millisecond Pulsars in the Globular Cluster 47 Tucanae}",
      journal = {\apj},
         year = 2006,
        month = aug,
       volume = {646},
       number = {2},
        pages = {1104-1115},
          doi = {10.1086/505133},
archivePrefix = {arXiv},
       eprint = {astro-ph/0604318},
 primaryClass = {astro-ph},
       adsurl = {https://ui.adsabs.harvard.edu/abs/2006ApJ...646.1104B}
}

@ARTICLE{2023MNRAS.518.1642Abbate,
       author = {{Abbate}, F. and {Possenti}, A. and {Ridolfi}, A. and {Venkatraman Krishnan}, V. and {Buchner}, S. and {Barr}, E.~D. and {Bailes}, M. and {Kramer}, M. and {Cameron}, A. and {Parthasarathy}, A. and {van Straten}, W. and {Chen}, W. and {Camilo}, F. and {Padmanabh}, P.~V. @ARTICLE{2018MNRAS.481..627A,
       author = {{Abbate}, F. and {Possenti}, A. and {Ridolfi}, A. and {Freire}, P.~C.~C. and {Camilo}, F. and {Manchester}, R.~N. and {D'Amico}, N.},
        title = "{Internal gas models and central black hole in 47 Tucanae using millisecond pulsars}",
      journal = {\mnras},
     keywords = {stars: kinematics and dynamics, pulsars: general, ISM: kinematics and dynamics, globular clusters: individual: 47 Tucanae, Astrophysics - High Energy Astrophysical Phenomena},
         year = 2018,
        month = nov,
       volume = {481},
       number = {1},
        pages = {627-638},
          doi = {10.1093/mnras/sty2298},
archivePrefix = {arXiv},
       eprint = {1808.06621},
 primaryClass = {astro-ph.HE},
       adsurl = {https://ui.adsabs.harvard.edu/abs/2018MNRAS.481..627A},
      adsnote = {Provided by the SAO/NASA Astrophysics Data System}
}

and {Mao}, S.~A. and {Freire}, P.~C.~C. and {Ransom}, S.~M. and {Vleeschower}, L. and {Geyer}, M. and {Zhang}, L.},
        title = "{A MeerKAT look at the polarization of 47 Tucanae pulsars: magnetic field implications}",
      journal = {\mnras},
         year = 2023,
        month = jan,
       volume = {518},
       number = {2},
        pages = {1642-1655},
          doi = {10.1093/mnras/stac3248},
archivePrefix = {arXiv},
       eprint = {2211.03815},
 primaryClass = {astro-ph.HE},
       adsurl = {https://ui.adsabs.harvard.edu/abs/2023MNRAS.518.1642A}
}

@ARTICLE{2018MNRAS.481..627Abbate,
       author = {{Abbate}, F. and {Possenti}, A. and {Ridolfi}, A. and {Freire}, P.~C.~C. and {Camilo}, F. and {Manchester}, R.~N. and {D'Amico}, N.},
        title = "{Internal gas models and central black hole in 47 Tucanae using millisecond pulsars}",
      journal = {\mnras},
         year = 2018,
        month = nov,
       volume = {481},
       number = {1},
        pages = {627-638},
          doi = {10.1093/mnras/sty2298},
archivePrefix = {arXiv},
       eprint = {1808.06621},
 primaryClass = {astro-ph.HE},
       adsurl = {https://ui.adsabs.harvard.edu/abs/2018MNRAS.481..627A}
}

@ARTICLE{2018MNRAS.478.1520Baumgardt,
       author = {{Baumgardt}, H. and {Hilker}, M.},
        title = "{A catalogue of masses, structural parameters, and velocity dispersion profiles of 112 Milky Way globular clusters}",
      journal = {\mnras},
         year = 2018,
        month = aug,
       volume = {478},
       number = {2},
        pages = {1520-1557},
          doi = {10.1093/mnras/sty1057},
archivePrefix = {arXiv},
       eprint = {1804.08359},
 primaryClass = {astro-ph.GA},
       adsurl = {https://ui.adsabs.harvard.edu/abs/2018MNRAS.478.1520B}
}

@ARTICLE{1975AJ.....80..809Hills,
       author = {{Hills}, J.~G.},
        title = "{Encounters between binary and single stars and their effect on the dynamical evolution of stellar systems.}",
      journal = {\aj},
         year = 1975,
        month = oct,
       volume = {80},
        pages = {809-825},
          doi = {10.1086/111815},
       adsurl = {https://ui.adsabs.harvard.edu/abs/1975AJ.....80..809H}
}

@ARTICLE{1995ApJS...99..609Sigurdsson,
       author = {{Sigurdsson}, Steinn and {Phinney}, E.~S.},
        title = "{Dynamics and Interactions of Binaries and Neutron Stars in Globular Clusters}",
      journal = {\apjs},
         year = 1995,
        month = aug,
       volume = {99},
        pages = {609},
          doi = {10.1086/192199},
archivePrefix = {arXiv},
       eprint = {astro-ph/9412078},
 primaryClass = {astro-ph},
       adsurl = {https://ui.adsabs.harvard.edu/abs/1995ApJS...99..609S}
}

@INPROCEEDINGS{1987IAUS..125..187Verbunt,
       author = {{Verbunt}, F. and {Hut}, P.},
        title = "{The Globular Cluster Population of X-Ray Binaries}",
    booktitle = {The Origin and Evolution of Neutron Stars},
         year = 1987,
       editor = {{Helfand}, D.~J. and {Huang}, J. -H.},
       series = {IAU Symposium},
       volume = {125},
        month = jan,
        pages = {187},
       adsurl = {https://ui.adsabs.harvard.edu/abs/1987IAUS..125..187V}
}

@ARTICLE{1975ApJ...199L.143Clark,
       author = {{Clark}, G.~W.},
        title = "{X-ray binaries in globular clusters.}",
      journal = {\apjl},
         year = 1975,
        month = aug,
       volume = {199},
        pages = {L143-L145},
          doi = {10.1086/181869},
       adsurl = {https://ui.adsabs.harvard.edu/abs/1975ApJ...199L.143C}
}

@ARTICLE{1982Natur.300..728Alpar,
       author = {{Alpar}, M.~A. and {Cheng}, A.~F. and {Ruderman}, M.~A. and {Shaham}, J.},
        title = "{A new class of radio pulsars}",
      journal = {\nat},
         year = 1982,
        month = dec,
       volume = {300},
       number = {5894},
        pages = {728-730},
          doi = {10.1038/300728a0},
       adsurl = {https://ui.adsabs.harvard.edu/abs/1982Natur.300..728A}
}

@ARTICLE{2013Natur.501..517Papitto,
       author = {{Papitto}, A. and {Ferrigno}, C. and {Bozzo}, E. and {Rea}, N. and {Pavan}, L. and {Burderi}, L. and {Burgay}, M. and {Campana}, S. and {di Salvo}, T. and {Falanga}, M. and {Filipovi{\'c}}, M.~D. and {Freire}, P.~C.~C. and {Hessels}, J.~W.~T. and {Possenti}, A. and {Ransom}, S.~M. and {Riggio}, A. and {Romano}, P. and {Sarkissian}, J.~M. and {Stairs}, I.~H. and {Stella}, L. and {Torres}, D.~F. and {Wieringa}, M.~H. and {Wong}, G.~F.},
        title = "{Swings between rotation and accretion power in a binary millisecond pulsar}",
      journal = {\nat},
         year = 2013,
        month = sep,
       volume = {501},
       number = {7468},
        pages = {517-520},
          doi = {10.1038/nature12470},
archivePrefix = {arXiv},
       eprint = {1305.3884},
 primaryClass = {astro-ph.HE},
       adsurl = {https://ui.adsabs.harvard.edu/abs/2013Natur.501..517P}
}

@ARTICLE{1982CSci...51.1096Radhakrishnan,
       author = {{Radhakrishnan}, V. and {Srinivasan}, G.},
        title = "{On the origin of the recently discovered ultra-rapid pulsar}",
      journal = {Current Science},
         year = 1982,
        month = dec,
       volume = {51},
        pages = {1096-1099},
       adsurl = {https://ui.adsabs.harvard.edu/abs/1982CSci...51.1096R}
}

@ARTICLE{1991PhR...203....1Bhattacharya,
       author = {{Bhattacharya}, D. and {van den Heuvel}, E.~P.~J.},
        title = "{Formation and evolution of binary and millisecond radio pulsars}",
      journal = {\physrep},
         year = 1991,
        month = jan,
       volume = {203},
       number = {1-2},
        pages = {1-124},
          doi = {10.1016/0370-1573(91)90064-S},
       adsurl = {https://ui.adsabs.harvard.edu/abs/1991PhR...203....1B}
}

@INCOLLECTION{2006csxs.book..623Tauris,
       author = {{Tauris}, T.~M. and {van den Heuvel}, E.~P.~J.},
        title = "{Formation and evolution of compact stellar X-ray sources}",
    booktitle = {Compact stellar X-ray sources},
         year = 2006,
       editor = {{Lewin}, Walter H.~G. and {van der Klis}, Michiel},
       volume = {39},
        pages = {623-665},
          doi = {10.48550/arXiv.astro-ph/0303456},
       adsurl = {https://ui.adsabs.harvard.edu/abs/2006csxs.book..623T}
}

@ARTICLE{2000ApJ...530L..93Tauris,
       author = {{Tauris}, Thomas M. and {van den Heuvel}, Edward P.~J. and {Savonije}, Gerrit J.},
        title = "{Formation of Millisecond Pulsars with Heavy White Dwarf Companions:Extreme Mass Transfer on Subthermal Timescales}",
      journal = {\apjl},
         year = 2000,
        month = feb,
       volume = {530},
       number = {2},
        pages = {L93-L96},
          doi = {10.1086/312496},
archivePrefix = {arXiv},
       eprint = {astro-ph/0001013},
 primaryClass = {astro-ph},
       adsurl = {https://ui.adsabs.harvard.edu/abs/2000ApJ...530L..93T}
}

@ARTICLE{2011MNRAS.416.2130Tauris,
       author = {{Tauris}, T.~M. and {Langer}, N. and {Kramer}, M.},
        title = "{Formation of millisecond pulsars with CO white dwarf companions - I. PSR J1614-2230: evidence for a neutron star born massive}",
      journal = {\mnras},
         year = 2011,
        month = sep,
       volume = {416},
       number = {3},
        pages = {2130-2142},
          doi = {10.1111/j.1365-2966.2011.19189.x},
archivePrefix = {arXiv},
       eprint = {1103.4996},
 primaryClass = {astro-ph.SR},
       adsurl = {https://ui.adsabs.harvard.edu/abs/2011MNRAS.416.2130T}
}

@ARTICLE{2012MNRAS.425.1601Tauris,
       author = {{Tauris}, T.~M. and {Langer}, N. and {Kramer}, M.},
        title = "{Formation of millisecond pulsars with CO white dwarf companions - II. Accretion, spin-up, true ages and comparison to MSPs with He white dwarf companions}",
      journal = {\mnras},
         year = 2012,
        month = sep,
       volume = {425},
       number = {3},
        pages = {1601-1627},
          doi = {10.1111/j.1365-2966.2012.21446.x},
archivePrefix = {arXiv},
       eprint = {1206.1862},
 primaryClass = {astro-ph.SR},
       adsurl = {https://ui.adsabs.harvard.edu/abs/2012MNRAS.425.1601T}
}

@ARTICLE{2012ApJ...745..109Lynch,
       author = {{Lynch}, Ryan S. and {Freire}, Paulo C.~C. and {Ransom}, Scott M. and {Jacoby}, Bryan A.},
        title = "{The Timing of Nine Globular Cluster Pulsars}",
      journal = {\apj},
         year = 2012,
        month = feb,
       volume = {745},
       number = {2},
          eid = {109},
        pages = {109},
          doi = {10.1088/0004-637X/745/2/109},
archivePrefix = {arXiv},
       eprint = {1112.2612},
 primaryClass = {astro-ph.HE},
       adsurl = {https://ui.adsabs.harvard.edu/abs/2012ApJ...745..109L}
}

@ARTICLE{2014A&A...561A..11Verbunt,
       author = {{Verbunt}, Frank and {Freire}, Paulo C.~C.},
        title = "{On the disruption of pulsar and X-ray binar ies in globular clusters}",
      journal = {\aap},
         year = 2014,
        month = jan,
       volume = {561},
          eid = {A11},
        pages = {A11},
          doi = {10.1051/0004-6361/201321177},
archivePrefix = {arXiv},
       eprint = {1310.4669},
 primaryClass = {astro-ph.SR},
       adsurl = {https://ui.adsabs.harvard.edu/abs/2014A&A...561A..11V}
}

@ARTICLE{2011PASA...28....1vanStraten,
       author = {{van Straten}, W. and {Bailes}, M.},
        title = "{DSPSR: Digital Signal Processing Software for Pulsar Astronomy}",
      journal = {\pasa},
         year = 2011,
        month = jan,
       volume = {28},
       number = {1},
        pages = {1-14},
          doi = {10.1071/AS10021},
archivePrefix = {arXiv},
       eprint = {1008.3973},
 primaryClass = {astro-ph.IM},
       adsurl = {https://ui.adsabs.harvard.edu/abs/2011PASA...28....1V}
}

@ARTICLE{2004PASA...21..302Hotan,
       author = {{Hotan}, A.~W. and {van Straten}, W. and {Manchester}, R.~N.},
        title = "{PSRCHIVE and PSRFITS: An Open Approach to Radio Pulsar Data Storage and Analysis}",
      journal = {\pasa},
         year = 2004,
        month = jan,
       volume = {21},
       number = {3},
        pages = {302-309},
          doi = {10.1071/AS04022},
archivePrefix = {arXiv},
       eprint = {astro-ph/0404549},
 primaryClass = {astro-ph},
       adsurl = {https://ui.adsabs.harvard.edu/abs/2004PASA...21..302H}
}

@ARTICLE{2022A&A...664A..27Ridolfi,
       author = {{Ridolfi}, A. and {Freire}, P.~C.~C. and {Gautam}, T. and {Ransom}, S.~M. and {Barr}, E.~D. and {Buchner}, S. and {Burgay}, M. and {Abbate}, F. and {Venkatraman Krishnan}, V. and {Vleeschower}, L. and {Possenti}, A. and {Stappers}, B.~W. and {Kramer}, M. and {Chen}, W. and {Padmanabh}, P.~V. and {Champion}, D.~J. and {Bailes}, M. and {Levin}, L. and {Keane}, E.~F. and {Breton}, R.~P. and {Bezuidenhout}, M. and {Grie{\ss}meier}, J. -M. and {K{\"u}nkel}, L. and {Men}, Y. and {Camilo}, F. and {Geyer}, M. and {Hugo}, B.~V. and {Jameson}, A. and {Parthasarathy}, A. and {Serylak}, M.},
        title = "{TRAPUM discovery of 13 new pulsars in NGC 1851 using MeerKAT}",
      journal = {\aap},
         year = 2022,
        month = aug,
       volume = {664},
          eid = {A27},
        pages = {A27},
          doi = {10.1051/0004-6361/202143006},
archivePrefix = {arXiv},
       eprint = {2203.12302},
 primaryClass = {astro-ph.HE},
       adsurl = {https://ui.adsabs.harvard.edu/abs/2022A&A...664A..27R}
}

@ARTICLE{1995ApJ...445L.133Rasio,
       author = {{Rasio}, Frederic A. and {Heggie}, Douglas C.},
        title = "{The Orbital Eccentricities of Binary Millisecond Pulsars in Globular Clusters}",
      journal = {\apjl},
         year = 1995,
        month = jun,
       volume = {445},
        pages = {L133},
          doi = {10.1086/187907},
archivePrefix = {arXiv},
       eprint = {astro-ph/9502105},
 primaryClass = {astro-ph},
       adsurl = {https://ui.adsabs.harvard.edu/abs/1995ApJ...445L.133R}
}

@software{2015ascl.soft09002Nice_tempo,
       author = {{Nice}, D. and {Demorest}, P. and {Stairs}, I. and {Manchester}, R. and {Taylor}, J. and {Peters}, W. and {Weisberg}, J. and {Irwin}, A. and {Wex}, N. and {Huang}, Y.},
        title = "{Tempo: Pulsar timing data analysis}",
 howpublished = {Astrophysics Source Code Library, record ascl:1509.002},
         year = 2015,
        month = sep,
          eid = {ascl:1509.002},
       adsurl = {https://ui.adsabs.harvard.edu/abs/2015ascl.soft09002N}
}

@ARTICLE{2016AJ....152...41Prsa,
       author = {{Pr{\v{s}}a}, Andrej and {Harmanec}, Petr and {Torres}, Guillermo and {Mamajek}, Eric and {Asplund}, Martin and {Capitaine}, Nicole and {Christensen-Dalsgaard}, J{\o}rgen and {Depagne}, {\'E}ric and {Haberreiter}, Margit and {Hekker}, Saskia and {Hilton}, James and {Kopp}, Greg and {Kostov}, Veselin and {Kurtz}, Donald W. and {Laskar}, Jacques and {Mason}, Brian D. and {Milone}, Eugene F. and {Montgomery}, Michele and {Richards}, Mercedes and {Schmutz}, Werner and {Schou}, Jesper and {Stewart}, Susan G.},
        title = "{Nominal Values for Selected Solar and Planetary Quantities: IAU 2015 Resolution B3}",
      journal = {\aj},
         year = 2016,
        month = aug,
       volume = {152},
       number = {2},
          eid = {41},
        pages = {41},
          doi = {10.3847/0004-6256/152/2/41},
archivePrefix = {arXiv},
       eprint = {1605.09788},
 primaryClass = {astro-ph.SR},
       adsurl = {https://ui.adsabs.harvard.edu/abs/2016AJ....152...41P}
}

@ARTICLE{1982ApJ...253..908Taylor,
       author = {{Taylor}, J.~H. and {Weisberg}, J.~M.},
        title = "{A new test of general relativity - Gravitational radiation and the binary pulsar PSR 1913+16}",
      journal = {\apj},
         year = 1982,
        month = feb,
       volume = {253},
        pages = {908-920},
          doi = {10.1086/159690},
       adsurl = {https://ui.adsabs.harvard.edu/abs/1982ApJ...253..908T}
}

@ARTICLE{2018MNRAS.476.4794Freire_dracula,
       author = {{Freire}, Paulo C.~C. and {Ridolfi}, Alessandro},
        title = "{An algorithm for determining the rotation count of pulsars}",
      journal = {\mnras},
         year = 2018,
        month = jun,
       volume = {476},
       number = {4},
        pages = {4794-4805},
          doi = {10.1093/mnras/sty524},
archivePrefix = {arXiv},
       eprint = {1802.07211},
 primaryClass = {astro-ph.IM},
       adsurl = {https://ui.adsabs.harvard.edu/abs/2018MNRAS.476.4794F}
}

@ARTICLE{2020PASA...37...12Hobbs,
       author = {{Hobbs}, George and {Manchester}, Richard N. and {Dunning}, Alex and {Jameson}, Andrew and {Roberts}, Paul and {George}, Daniel and {Green}, J.~A. and {Tuthill}, John and {Toomey}, Lawrence and {Kaczmarek}, Jane F. and {Mader}, Stacy and {Marquarding}, Malte and {Ahmed}, Azeem and {Amy}, Shaun W. and {Bailes}, Matthew and {Beresford}, Ron and {Bhat}, N.~D.~R. and {Bock}, Douglas C. -J. and {Bourne}, Michael and {Bowen}, Mark and {Brothers}, Michael and {Cameron}, Andrew D. and {Carretti}, Ettore and {Carter}, Nick and {Castillo}, Santy and {Chekkala}, Raji and {Cheng}, Wan and {Chung}, Yoon and {Craig}, Daniel A. and {Dai}, Shi and {Dawson}, Joanne and {Dempsey}, James and {Doherty}, Paul and {Dong}, Bin and {Edwards}, Philip and {Ergesh}, Tuohutinuer and {Gao}, Xuyang and {Han}, JinLin and {Hayman}, Douglas and {Indermuehle}, Balthasar and {Jeganathan}, Kanapathippillai and {Johnston}, Simon and {Kanoniuk}, Henry and {Kesteven}, Michael and {Kramer}, Michael and {Leach}, Mark and {Mcintyre}, Vince and {Moss}, Vanessa and {Os{\l}owski}, Stefan and {Phillips}, Chris and {Pope}, Nathan and {Preisig}, Brett and {Price}, Daniel and {Reeves}, Ken and {Reilly}, Les and {Reynolds}, John and {Robishaw}, Tim and {Roush}, Peter and {Ruckley}, Tim and {Sadler}, Elaine and {Sarkissian}, John and {Severs}, Sean and {Shannon}, Ryan and {Smart}, Ken and {Smith}, Malcolm and {Smith}, Stephanie and {Sobey}, Charlotte and {Staveley-Smith}, Lister and {Tzioumis}, Anastasios and {van Straten}, Willem and {Wang}, Nina and {Wen}, Linqing and {Whiting}, Matthew},
        title = "{An ultra-wide bandwidth (704 to 4 032 MHz) receiver for the Parkes radio telescope}",
      journal = {\pasa},
         year = 2020,
        month = apr,
       volume = {37},
          eid = {e012},
        pages = {e012},
          doi = {10.1017/pasa.2020.2},
archivePrefix = {arXiv},
       eprint = {1911.00656},
 primaryClass = {astro-ph.IM},
       adsurl = {https://ui.adsabs.harvard.edu/abs/2020PASA...37...12H}
}

@ARTICLE{2012AJ....143...50Woodley,
       author = {{Woodley}, K.~A. and {Goldsbury}, R. and {Kalirai}, J.~S. and {Richer}, H.~B. and {Tremblay}, P. -E. and {Anderson}, J. and {Bergeron}, P. and {Dotter}, A. and {Esteves}, L. and {Fahlman}, G.~G. and {Hansen}, B.~M.~S. and {Heyl}, J. and {Hurley}, J. and {Rich}, R.~M. and {Shara}, M.~M. and {Stetson}, P.~B.},
        title = "{The Spectral Energy Distributions of White Dwarfs in 47 Tucanae: The Distance to the Cluster}",
      journal = {\aj},
         year = 2012,
        month = feb,
       volume = {143},
       number = {2},
          eid = {50},
        pages = {50},
          doi = {10.1088/0004-6256/143/2/50},
archivePrefix = {arXiv},
       eprint = {1112.1425},
 primaryClass = {astro-ph.GA},
       adsurl = {https://ui.adsabs.harvard.edu/abs/2012AJ....143...50W}
}

@ARTICLE{2017ApJ...851L..29Martinez,
       author = {{Martinez}, J.~G. and {Stovall}, K. and {Freire}, P.~C.~C. and {Deneva}, J.~S. and {Tauris}, T.~M. and {Ridolfi}, A. and {Wex}, N. and {Jenet}, F.~A. and {McLaughlin}, M.~A. and {Bagchi}, M.},
        title = "{Pulsar J1411+2551: A Low-mass Double Neutron Star System}",
      journal = {\apjl},
         year = 2017,
        month = dec,
       volume = {851},
       number = {2},
          eid = {L29},
        pages = {L29},
          doi = {10.3847/2041-8213/aa9d87},
archivePrefix = {arXiv},
       eprint = {1711.09804},
 primaryClass = {astro-ph.HE},
       adsurl = {https://ui.adsabs.harvard.edu/abs/2017ApJ...851L..29M}
}

@ARTICLE{2013ApJ...774..151Miocchi,
       author = {{Miocchi}, P. and {Lanzoni}, B. and {Ferraro}, F.~R. and {Dalessandro}, E. and {Vesperini}, E. and {Pasquato}, M. and {Beccari}, G. and {Pallanca}, C. and {Sanna}, N.},
        title = "{Star Count Density Profiles and Structural Parameters of 26 Galactic Globular Clusters}",
      journal = {\apj},
         year = 2013,
        month = sep,
       volume = {774},
       number = {2},
          eid = {151},
        pages = {151},
          doi = {10.1088/0004-637X/774/2/151},
archivePrefix = {arXiv},
       eprint = {1307.6035},
 primaryClass = {astro-ph.GA},
       adsurl = {https://ui.adsabs.harvard.edu/abs/2013ApJ...774..151M}
}

@ARTICLE{2018ApJ...854L..22Stovall,
       author = {{Stovall}, K. and {Freire}, P.~C.~C. and {Chatterjee}, S. and {Demorest}, P.~B. and {Lorimer}, D.~R. and {McLaughlin}, M.~A. and {Pol}, N. and {van Leeuwen}, J. and {Wharton}, R.~S. and {Allen}, B. and {Boyce}, M. and {Brazier}, A. and {Caballero}, K. and {Camilo}, F. and {Camuccio}, R. and {Cordes}, J.~M. and {Crawford}, F. and {Deneva}, J.~S. and {Ferdman}, R.~D. and {Hessels}, J.~W.~T. and {Jenet}, F.~A. and {Kaspi}, V.~M. and {Knispel}, B. and {Lazarus}, P. and {Lynch}, R. and {Parent}, E. and {Patel}, C. and {Pleunis}, Z. and {Ransom}, S.~M. and {Scholz}, P. and {Seymour}, A. and {Siemens}, X. and {Stairs}, I.~H. and {Swiggum}, J. and {Zhu}, W.~W.},
        title = "{PALFA Discovery of a Highly Relativistic Double Neutron Star Binary}",
      journal = {\apjl},
         year = 2018,
        month = feb,
       volume = {854},
       number = {2},
          eid = {L22},
        pages = {L22},
          doi = {10.3847/2041-8213/aaad06},
archivePrefix = {arXiv},
       eprint = {1802.01707},
 primaryClass = {astro-ph.HE},
       adsurl = {https://ui.adsabs.harvard.edu/abs/2018ApJ...854L..22S}
}

@ARTICLE{Bailes+2020,
       author = {{Bailes}, M. and {Jameson}, A. and {Abbate}, F. and {Barr}, E.~D. and {Bhat}, N.~D.~R. and {Bondonneau}, L. and {Burgay}, M. and {Buchner}, S.~J. and {Camilo}, F. and {Champion}, D.~J. and {Cognard}, I. and {Demorest}, P.~B. and {Freire}, P.~C.~C. and {Gautam}, T. and {Geyer}, M. and {Griessmeier}, J. -M. and {Guillemot}, L. and {Hu}, H. and {Jankowski}, F. and {Johnston}, S. and {Karastergiou}, A. and {Karuppusamy}, R. and {Kaur}, D. and {Keith}, M.~J. and {Kramer}, M. and {van Leeuwen}, J. and {Lower}, M.~E. and {Maan}, Y. and {McLaughlin}, M.~A. and {Meyers}, B.~W. and {Os{\l}owski}, S. and {Oswald}, L.~S. and {Parthasarathy}, A. and {Pennucci}, T. and {Posselt}, B. and {Possenti}, A. and {Ransom}, S.~M. and {Reardon}, D.~J. and {Ridolfi}, A. and {Schollar}, C.~T.~G. and {Serylak}, M. and {Shaifullah}, G. and {Shamohammadi}, M. and {Shannon}, R.~M. and {Sobey}, C. and {Song}, X. and {Spiewak}, R. and {Stairs}, I.~H. and {Stappers}, B.~W. and {van Straten}, W. and {Szary}, A. and {Theureau}, G. and {Venkatraman Krishnan}, V. and {Weltevrede}, P. and {Wex}, N. and {Abbott}, T.~D. and {Adams}, G.~B. and {Burger}, J.~P. and {Gamatham}, R.~R.~G. and {Gouws}, M. and {Horn}, D.~M. and {Hugo}, B. and {Joubert}, A.~F. and {Manley}, J.~R. and {McAlpine}, K. and {Passmoor}, S.~S. and {Peens-Hough}, A. and {Ramudzuli}, Z.~R. and {Rust}, A. and {Salie}, S. and {Schwardt}, L.~C. and {Siebrits}, R. and {Van Tonder}, G. and {Van Tonder}, V. and {Welz}, M.~G.},
        title = "{The MeerKAT telescope as a pulsar facility: System verification and early science results from MeerTime}",
      journal = {\pasa},
         year = 2020,
        month = jul,
       volume = {37},
          eid = {e028},
        pages = {e028},
          doi = {10.1017/pasa.2020.19},
archivePrefix = {arXiv},
       eprint = {2005.14366},
 primaryClass = {astro-ph.IM},
       adsurl = {https://ui.adsabs.harvard.edu/abs/2020PASA...37...28B}
}

@ARTICLE{2017ApJ...846..170Tauris,
       author = {{Tauris}, T.~M. and {Kramer}, M. and {Freire}, P.~C.~C. and {Wex}, N. and {Janka}, H. -T. and {Langer}, N. and {Podsiadlowski}, Ph. and {Bozzo}, E. and {Chaty}, S. and {Kruckow}, M.~U. and {van den Heuvel}, E.~P.~J. and {Antoniadis}, J. and {Breton}, R.~P. and {Champion}, D.~J.},
        title = "{Formation of Double Neutron Star Systems}",
      journal = {\apj},
         year = 2017,
        month = sep,
       volume = {846},
       number = {2},
          eid = {170},
        pages = {170},
          doi = {10.3847/1538-4357/aa7e89},
archivePrefix = {arXiv},
       eprint = {1706.09438},
 primaryClass = {astro-ph.HE},
       adsurl = {https://ui.adsabs.harvard.edu/abs/2017ApJ...846..170T}
}

@ARTICLE{1991Natur.352..219_Manchester,
       author = {{Manchester}, R.~N. and {Lyne}, A.~G. and {Robinson}, C. and {D'Amico}, N. and {Bailes}, M. and {Lim}, J.},
        title = "{Discovery of ten millisecond pulsars in the globular cluster 47 Tucanae}",
      journal = {\nat},
         year = 1991,
        month = jul,
       volume = {352},
       number = {6332},
        pages = {219-221},
          doi = {10.1038/352219a0},
       adsurl = {https://ui.adsabs.harvard.edu/abs/1991Natur.352..219M}
}

@ARTICLE{2024A&A...686A.166_Padmanabh,
       author = {{Padmanabh}, P.~V. and {Ransom}, S.~M. and {Freire}, P.~C.~C. and {Ridolfi}, A. and {Taylor}, J.~D. and {Choza}, C. and {Clark}, C.~J. and {Abbate}, F. and {Bailes}, M. and {Barr}, E.~D. and {Buchner}, S. and {Burgay}, M. and {DeCesar}, M.~E. and {Chen}, W. and {Corongiu}, A. and {Champion}, D.~J. and {Dutta}, A. and {Geyer}, M. and {Hessels}, J.~W.~T. and {Kramer}, M. and {Possenti}, A. and {Stairs}, I.~H. and {Stappers}, B.~W. and {Venkatraman Krishnan}, V. and {Vleeschower}, L. and {Zhang}, L.},
        title = "{Discovery and timing of ten new millisecond pulsars in the globular cluster Terzan 5}",
      journal = {\aap},
         year = 2024,
        month = jun,
       volume = {686},
          eid = {A166},
        pages = {A166},
          doi = {10.1051/0004-6361/202449303},
archivePrefix = {arXiv},
       eprint = {2403.17799},
 primaryClass = {astro-ph.HE},
       adsurl = {https://ui.adsabs.harvard.edu/abs/2024A&A...686A.166P}
}

@ARTICLE{2005AJ....129.1993Manchester,
       author = {{Manchester}, R.~N. and {Hobbs}, G.~B. and {Teoh}, A. and {Hobbs}, M.},
        title = "{The Australia Telescope National Facility Pulsar Catalogue}",
      journal = {\aj},
         year = 2005,
        month = apr,
       volume = {129},
       number = {4},
        pages = {1993-2006},
          doi = {10.1086/428488},
archivePrefix = {arXiv},
       eprint = {astro-ph/0412641},
 primaryClass = {astro-ph},
       adsurl = {https://ui.adsabs.harvard.edu/abs/2005AJ....129.1993M}
}

@article{Robertson_1938_twobodyprobinGR,
 ISSN = {0003486X, 19398980},
 URL = {http://www.jstor.org/stable/1968715},
 author = {H. P. Robertson},
 journal = {Annals of Mathematics},
 number = {1},
 pages = {101--104},
 publisher = {[Annals of Mathematics, Trustees of Princeton University on Behalf of the Annals of Mathematics, Mathematics Department, Princeton University]},
 title = {Note on the Preceding Paper: The Two Body Problem in General Relativity},
 urldate = {2025-09-11},
 volume = {39},
 year = {1938}
}

@ARTICLE{2017MNRAS.468..645Brogaard,
       author = {{Brogaard}, K. and {VandenBerg}, D.~A. and {Bedin}, L.~R. and {Milone}, A.~P. and {Thygesen}, A. and {Grundahl}, F.},
        title = "{The age of 47 Tuc from self-consistent isochrone fits to colour-magnitude diagrams and the eclipsing member V69}",
      journal = {\mnras},
         year = 2017,
        month = jun,
       volume = {468},
       number = {1},
        pages = {645-661},
          doi = {10.1093/mnras/stx378},
archivePrefix = {arXiv},
       eprint = {1702.03421},
 primaryClass = {astro-ph.SR},
       adsurl = {https://ui.adsabs.harvard.edu/abs/2017MNRAS.468..645B}
}

@ARTICLE{2003A&A...408..Gratton,
       author = {{Gratton}, R.~G. and {Bragaglia}, A. and {Carretta}, E. and {Clementini}, G. and {Desidera}, S. and {Grundahl}, F. and {Lucatello}, S.},
        title = "{Distances and ages of NGC 6397, NGC 6752 and 47 Tuc}",
      journal = {\aap},
         year = 2003,
        month = sep,
       volume = {408},
        pages = {529-543},
          doi = {10.1051/0004-6361:20031003},
archivePrefix = {arXiv},
       eprint = {astro-ph/0307016},
 primaryClass = {astro-ph},
       adsurl = {https://ui.adsabs.harvard.edu/abs/2003A&A...408..529G}
}

@ARTICLE{2020MNRAS.492.Thompson,
       author = {{Thompson}, I.~B. and {Udalski}, A. and {Dotter}, A. and {Rozyczka}, M. and {Schwarzenberg-Czerny}, A. and {Pych}, W. and {Beletsky}, Y. and {Burley}, G.~S. and {Marshall}, J.~L. and {McWilliam}, A. and {Morrell}, N. and {Osip}, D. and {Monson}, A. and {Persson}, S.~E. and {Szyma{\'n}ski}, M.~K. and {Soszy{\'n}ski}, I. and {Poleski}, R. and {Ulaczyk}, K. and {Wyrzykowski}, {\L}. and {Koz{\l}owski}, S. and {Mr{\'o}z}, P. and {Pietrukowicz}, P. and {Skowron}, J.},
        title = "{The Cluster AgeS Experiment (CASE) - VIII. Age and distance of the Globular Cluster 47 Tuc from the analysis of two detached eclipsing binaries}",
      journal = {\mnras},
         year = 2020,
        month = mar,
       volume = {492},
       number = {3},
        pages = {4254-4267},
          doi = {10.1093/mnras/staa032},
archivePrefix = {arXiv},
       eprint = {2001.01481},
 primaryClass = {astro-ph.SR},
       adsurl = {https://ui.adsabs.harvard.edu/abs/2020MNRAS.492.4254T}
}

@ARTICLE{2025ApJS..279...51Lian,
       author = {{Lian}, Yujie and {Pan}, Zhichen and {Zhang}, Haiyan and {Cao}, Shuo and {Freire}, P.~C.~C. and {Qian}, Lei and {Eatough}, Ralph P. and {Shao}, Lijing and {Ransom}, Scott M. and {Lorimer}, Duncan R. and {Yin}, Dejiang and {Dai}, Yinfeng and {Liu}, Kuo and {Wang}, Lin and {Wang}, Yujie and {Zhang}, Zhongli and {Feng}, Zhonghua and {Li}, Baoda and {Li}, Minghui and {Liu}, Tong and {Li}, Yaowei and {Peng}, Bo and {Pan}, Yu and {Wu}, Yuxiao and {Zhang}, Liyun and {Zhang}, Xingnan and {Jiang}, Peng},
        title = "{The FAST Globular Cluster Pulsar Survey (GC FANS)}",
      journal = {\apjs},
         year = 2025,
        month = aug,
       volume = {279},
       number = {2},
          eid = {51},
        pages = {51},
          doi = {10.3847/1538-4365/ade4ba},
archivePrefix = {arXiv},
       eprint = {2506.07970},
 primaryClass = {astro-ph.HE},
       adsurl = {https://ui.adsabs.harvard.edu/abs/2025ApJS..279...51L}
}

@PHDTHESIS{Ransom2001,
   author = {{Ransom}, S.~M.},
    title = "{New search techniques for binary pulsars}",
   school = {Harvard University},
     year = 2001,
   adsurl = {http://adsabs.harvard.edu/abs/2001PhDT.......123R}
}

@ARTICLE{Staveley-Smith+1996,
   author = {{Staveley-Smith}, L. and {Wilson}, W.~E. and {Bird}, T.~S. and 
	{Disney}, M.~J. and {Ekers}, R.~D. and {Freeman}, K.~C. and 
	{Haynes}, R.~F. and {Sinclair}, M.~W. and {Vaile}, R.~A. and 
	{Webster}, R.~L. and {Wright}, A.~E.},
    title = "{The Parkes 21 CM multibeam receiver}",
  journal = {\pasa},
     year = 1996,
    month = nov,
   volume = 13,
    pages = {243-248},
   adsurl = {http://adsabs.harvard.edu/abs/1996PASA...13..243S}
}

@ARTICLE{2025A&A...704A.153Meng,
       author = {{Meng}, Lingqi and {Freire}, Paulo C.~C. and {Stovall}, Kevin and {Wex}, Norbert and {Miao}, Xueli and {Zhu}, Weiwei and {Kramer}, Michael and {Cordes}, James M. and {Hu}, Huanchen and {Jiang}, Jinchen and {Parent}, Emilie and {Shao}, Lijing and {Stairs}, Ingrid H. and {Xue}, Mengyao and {Brazier}, Adam and {Camilo}, Fernando and {Champion}, David J. and {Chatterjee}, Shami and {Crawford}, Fronefield and {Fang}, Ziyao and {Fu}, Qiuyang and {Guo}, Yanjun and {Hessels}, Jason W.~T. and {MacLaughlin}, Maura and {Miao}, Chenchen and {Niu}, Jiarui and {Wu}, Ziwei and {Yao}, Jumei and {Yuan}, Mao and {Yue}, Youlin and {Zhang}, Chengmin},
        title = "{The double neutron star PSR J1946+2052: I. Masses and tests of general relativity}",
      journal = {\aap},
         year = 2025,
        month = dec,
       volume = {704},
          eid = {A153},
        pages = {A153},
          doi = {10.1051/0004-6361/202555689},
archivePrefix = {arXiv},
       eprint = {2510.12506},
 primaryClass = {astro-ph.HE},
       adsurl = {https://ui.adsabs.harvard.edu/abs/2025A&A...704A.153M}
}

\end{document}